\documentclass[amssymb,showpacs,twocolumn]{revtex4}
\begin{document}

\newtheorem{theorem}{Theorem}
\newtheorem{corollary}{Corollary}

\newcommand{\dbar}{d \!\!\:\bar{}\,}
\newcommand{\Id}{I\!\!I}
\newcommand{\R}{\mathbb R}
\newcommand{\OP}{p}
\newcommand{\PP}{{\underline{p}_1}}
\newcommand{\BB}{I\!\!B}
\newcommand{\lie}{{\cal L}}
\newcommand{\qed}{\hfill \fbox{}}
\newcommand{\h}{\mbox{{\em l}}\!\hspace{0.1mm}\mbox{{\em h}}}
\newcommand{\B}{{\cal B}}
\newcommand{\C}{{\cal C}}

\newcommand{\N}{\mathbb{N}}
\newcommand{\un}{\underline}
\newcommand{\w}{\omega}
\newcommand{\W}{\langle \w \rangle}

\newcommand{\Stz}{{\cal S}}
\newcommand{\Scl}{S_{\mbox{\footnotesize{cl}}}}
\newcommand{\psicl}{\psi_{\mbox{\footnotesize{cl}}}}
\newcommand{\F}{{\cal F}}

\title{Strongly hyperbolic second order Einstein's evolution equations}

\author{Gabriel Nagy}
\email{gnagy@math.ucsd.edu}
\affiliation{Visiting scholar, Department of Mathematics, \\
University of California at San Diego, \\
9500 Gilman Drive, La Jolla, California 92093-0112, USA}

\author{Omar E. Ortiz}
\email{ortiz@fis.uncor.edu}
\author{Oscar A. Reula}
\email{reula@fis.uncor.edu}
\affiliation{Facultad de Matem\'atica Astronom\'{\i}a y F\'{\i}sica, \\
Universidad Nacional de C\'ordoba, \\
Ciudad Universitaria, 5000 C\'ordoba, Argentina}

\date{August 11, 2004}

\begin{abstract}
  BSSN-type evolution equations are discussed. The name refers to the
  Baumgarte, Shapiro, Shibata, and Nakamura version of the Einstein
  evolution equations, without introducing the conformal-traceless
  decomposition but keeping the three connection functions and
  including a densitized lapse.  It is proved that a
  pseudo-differential first order reduction of these equations is
  strongly hyperbolic. In the same way, densitized
  Arnowitt-Deser-Misner evolution equations are found to be weakly
  hyperbolic. In both cases, the positive densitized lapse function
  and the spacelike shift vector are arbitrary given fields. This
  first order pseudodifferential reduction adds no extra equations
  to the system and so no extra constraints.
\end{abstract}


\maketitle

\section{Introduction}

Einstein's equation determines geometries; hence its solutions are
equivalent classes under space-time diffeomorphisms of metric tensors.
It is this invariance, however, which imposes a particular aftermath
on every initial value formulation for Einstein's equation. The
geometrical equation must be first converted into a system having a
well posed Cauchy problem, and so without the diffeomorphism
invariance. A preferred foliation of spacelike hypersurfaces on the
space-time is usually introduced in order that adapted coordinates
break this invariance.  Einstein's equation is then decomposed into
constraint equations on the foliation hypersurfaces and evolution
equations. While the constraints are uniquely determined by this
procedure, the evolution equations are not. Some of these evolution
equations turn out to be hyperbolic. This is in accordance with a main
aspect of general relativity, that of causal propagation of the
gravitational field.

Hyperbolicity refers to algebraic conditions on the principal part of
the equations which imply well posedness for the Cauchy problem, that
is, the existence of a unique continuous map between solutions and
initial data. There are several notions of hyperbolicity, which are
related to different algebraic conditions. Some notions imply well
posedness for the Cauchy problem in constant coefficient equations but
not in more general systems, such as quasilinear equations. See
\cite{oR98,hFaR00} for reviews intended for researchers on general
relativity. Regarding quasilinear systems, strong hyperbolicity is one
of the more general notions of hyperbolicity that implies well
posedness for the Cauchy problem. The proof involves
pseudodifferential analysis \cite{Taylor81,Taylor91}. Symmetric
hyperbolic systems are a particular case of strongly hyperbolic
systems where well posedness can be proved without using
pseudodifferential techniques. Several equations from physics can be
cast into this symmetric hyperbolic form \cite{rG96}.  Finally, weak
hyperbolicity is a less rigid notion than strong hyperbolicity but it
does not imply well posedness for quasilinear equations.

A definition of strong hyperbolicity for pseu\-do\-dif\-fer\-en\-tial
first order systems is introduced in Sec. \ref{S:wp}. Differential
first order strongly hyperbolic systems known in the literature are
included. A reason for this definition is that it incorporates
precisely the hypothesis needed for proving well posedness. The proof
involves standard pseudodifferential techniques. If an $m$-order
differential system has a first order, differential or
pseudodifferential, strongly hyperbolic reduction, then it is well
posed. See the end of this section for an example of a first order
pseudodifferential reduction of the wave equation. Also see Appendix
\ref{e} for a brief and self-contained introduction to the subject of
pseudodifferential operators.

Although strongly hyperbolic systems are at the core of the various
proofs of well posedness for the Cauchy problem in general relativity,
they have played, until recently, no similar role in numerical
relativity \cite{lL01}. Finite difference schemes have been
implemented for non-strongly-hyperbolic equations. However, Lax's
equivalence theorem does not hold in these situations
\cite{Richtmyer67}. It has been shown that discretization schemes
standard in numerical relativity are not convergent when applied to
weakly hyperbolic and ill posed systems \cite{gC-etal02a,gC-etal02b}.
As more complicated situations are studied numerically, the interest
in strongly hyperbolic reductions of Einstein's equation is
increasing. There is also much experience and a vast literature in
numerical schemes based on well posed formulations coming from
inviscid hydrodynamics \cite{Kreiss89,Gustafsson95}. This experience
can be transferred into numerical relativity when strongly hyperbolic
reductions are used.

Early numerical schemes to solve Einstein's equation were based on
variants of the Arnowitt-Deser-Misner (ADM) decomposition
\cite{ADM62}. Only recently has it been proven that a first order
differential reduction of the ADM evolution equations is weakly
hyperbolic \cite{lKmSsT01}. This is the reason for some of the
instabilities observed in ADM-based numerical schemes
\cite{tBsS98,gC-etal02a,gC-etal02b}. In the first part of this work
the ADM evolution equations are reviewed.  A densitized lapse function
is introduced, with the density exponent held as a free parameter. The
evolution equations are reduced to first order in a pseudodifferential
way. It is found that the resulting system is weakly hyperbolic for
every prescription of a positive density lapse and a spacelike shift
vector. This is summarized in Theorem \ref{T:ADMwh}.
pseudodifferential techniques and mode decomposition are at the core
of a proof directed to computing the eigenvalues and eigenvectors of the
principal symbol of the evolution equations, and to check that the
eigenvectors do not span the whole eigenspace. The mode decomposition
also helps to understand why the addition of the Hamiltonian
constraint into the system does not produce a strongly hyperbolic
system.

Baumgarte-Shapiro-Shibata-Nakamura-(BSSN-)type systems are introduced
in the second part of this work. They are essentially the densitized
ADM evolution equations where some combination of connection
coefficients of the three-metric is introduced as a new variable. The
mode decomposition of the densitized ADM evolution equations and
previous work on the linearized ADM equations \cite{hKoO02} suggest
the introduction of this variable.  It turns out to be related with
the variable $\tilde \Gamma^i$ of the BSSN system, defined by Eq. (21)
in \cite{tBsS98}. (See also \cite{mStN95}.) Similar variables have
been introduced in
\cite{cBjM92,cB-etal95,aA-etal99,cB-etal02,cB-etal03}. Their evolution
equation is obtained, as in the BSSN system, from commuting
derivatives and then adding the momentum constraint. It is shown here
that the addition of the momentum constraint transforms a weakly
hyperbolic system into a strongly hyperbolic one. This is the main
result of this work, and it is presented in Theorem \ref{T:BSSNsth}.
The first part of the proof follows the previous one for the
densitized ADM evolution equations. Once the eigenvectors are
computed, and it is verified that they do span the whole eigenspace,
the proof continues with the construction of the symmetrizer. This
construction is carried out with the eigenvectors. Finally, the
smooth properties of the symmetrizer are verified.

The hyperbolicity of a family of BSSN-type evolution equations has
previously been studied with a different technique \cite{oS-etal02}.
The equations were reduced to a differential system of first order in
time and in space derivatives. The lapse was densitized, the
eigenvalues and eigenvectors of the principal symbol were computed,
and it was verified that the latter do span their eigenspace. A
smooth symmetrizer was computed for a subfamily of systems, showing
strong hyperbolicity in this case.

All notions of hyperbolicity mentioned here require rewriting the
evolution equations as a first order system. This can be done in a
differential or pseudodifferential way. Some pseudodifferential
reductions to first order have the advantage that no extra equations
are added into the system, so there are no extra constraints. This
reduces the algebra needed to compute the symmetrizer. These
techniques are well known in the field of pseu\-do-dif\-fer\-en\-tial
calculus. They were first used in general relativity in \cite{hKoO02},
where linearized ADM evolution equations were proved to be weakly
hyperbolic. The example below presents the wave equation as a
toy model to understand how the pseudodifferential first order
reduction works. See Appendix \ref{e}, and also Sec. 5.3 in
\cite{Taylor91}, for other possible first order reductions.  Consider
the wave equation on $\R^4$ for a function $h$, written as a first
order system in time, in appropriate coordinates, that is,
\[
\partial_t h = k,
\quad
\partial_t k = \Delta h,
\]
with $t\in [0,\infty)$, $x^i \in \R^3$, and $\Delta = \delta^{ij}
\partial_i\partial_j$ the flat Laplacian.  Here $\delta^{ij} =
\mbox{diag}(1,1,1)$. Fourier transform the system in $x^i$,
\begin{equation}\label{tm}
\partial_t \hat h = \hat k,
\quad
\partial_t \hat k = - |\omega|^2 \hat h,
\end{equation}
where $\hat h(t,\omega_i)$ is the Fourier transform of $h(t,x^i)$ in
the space variables, as defined in Appendix \ref{S:fs}. The function
$\hat k$ is defined in an analogous way, and $|\omega|^2 =
\delta^{ij}\omega_i\omega_j$. The key step is to rewrite Eqs.
(\ref{tm}) as a first order system by introducing the unknown $\hat
\ell := i |\omega| \hat h$, where $i |\omega|$ is the symbol of the
pseudodifferential operator square root of the Laplacian. One gets
\[
\partial_t \hat \ell = i |\omega| \hat k,
\quad
\partial_t \hat k = i |\omega| \hat \ell.
\]
The system so obtained is a reduction to first order of the original
first order in time wave equation. Notice that there is no increase in
the number of unknowns, just a replacement of $\hat h$ by $\hat \ell$,
and correspondingly no extra constraints are introduced. For the wave
equation the result is a symmetric hyperbolic system. In the case of
the ADM equations the resulting system is weakly hyperbolic, with or
without densitizing the lapse function, while for the BSSN-type
equations one gets a strongly hyperbolic system.

In Sec. \ref{S:wp} a precise definition for well posedness is
introduced for first order quasilinear pseudodifferential systems.
Strongly hyperbolic systems are also defined and the main theorem
asserting well posedness for these systems is reviewed.  Section
\ref{s:ADM} is dedicated to reviewing the densitized ADM equations.
The main result here is Theorem \ref{T:ADMwh}, asserting that the
resulting evolution equations are weakly hyperbolic. The role of
adding the Hamiltonian constraint is briefly discussed. Section
\ref{s:BSSN} is dedicated to introducing the BSSN-type system
modifying the densitized ADM equations. The main result of this work,
Theorem \ref{T:BSSNsth}, asserts that this BSSN-type system is
strongly hyperbolic for some choices of the free parameters. The key
point is the introduction of the momentum constraint into the
evolution equations.  Section \ref{S:D} summarizes these results
briefly. Appendix \ref{e} is an introduction to pseudodifferential
calculus. It summarizes the main ideas and highlights the main
results. It is intended for physicists interested in learning the
subject. It provides all the background knowledge to follow the
calculations presented in this work. A summary of this type of
pseudodifferential calculus was not found by the authors in the
specialized literature.

\section{Well posedness}
\label{S:wp}

Hadamard first introduced the concept of well posedness for a Cauchy
problem. It essentially says that a well posed problem should have a
solution, that this solution should be unique, and that it should
depend continuously on the data of the problem. The first two
requirements are clear, but the last one needs additional
specifications. First, there is no unique way to prescribe this notion
of continuity.  Although a topological space is all that is needed to
introduce it, Banach spaces are present in most definitions of well
posedness. This refinement simplifies the analysis while still
including a large class of problems. Second, as nonlinear systems
lack, in general, global in time solutions, one can at most expect a
local in time notion of well posedness. Discussions on well posedness
can be found in \cite{Gustafsson95,Kreiss89} and
\cite{Taylor81,Taylor91}. See \cite{oR98} for a summary. This section
is dedicated to reviewing the minimum set of definitions and results
on well posedness for quasilinear pseudodifferential strongly
hyperbolic systems which are needed to describe the equations coming
from general relativity. It assumes that the reader is acquainted with
the notions from functional analysis and pseudodifferential calculus
given in Appendix \ref{e}.

Consider the Cauchy problem for a quasilinear first order
pseudodifferential system
\begin{equation}
\label{pdoc}
\partial_t u = p(t,x,u,\partial_x)u,\quad
u|_{t=0} = f,
\end{equation}
where $u$, $f$ are $m$-dimensional vector valued functions, $m\geq 1$,
and $x$ represents Cartesian coordinates in $\R^n$, $n\geq 1$.  Here
$p(t,x,v,\partial_x)$ is a smooth family of pseudodifferential
operators in $\psicl^1$, parametrized by $t\in \R^{+}$ and $v\in
\R^m$. Let $\un p(t,x,v,\w)$ and $\un p_1(t,x,v,\w)$ be their symbols
and principal symbols, respectively.

If $p$ is a differential operator with analytic coefficients, then the
Cauchy-Kowalewski theorem asserts that there exists a unique solution
for every analytic data $f$. However, solutions corresponding to
smooth data behave very differently depending on the type of operator
$p$. For example, write the flat Laplace equation in $\R^{n+1}$ and
the flat wave equation in $\R^{n+1}$ as first order systems in the
form $\partial_t u = A^i\partial_i u$. The matrices $A^i$ are
skew symmetric for the Laplacian, and symmetric for the wave operator.
Therefore, solutions of the form $u(t,x) = \hat u(t) e^{i\w\cdot x}$
for the corresponding Cauchy problems behave very differently at the high
frequency limit. The solutions of the Cauchy problem for the Laplace
equation diverge in the limit $|\w|\to\infty$, while the solutions of
the wave equation do not diverge in that limit.  (See footnote
\footnote{An explicit example in $\R^2$, presented by Hadamard in
  \cite{Hadamard23}, may clarify this. Consider the functions
$$
v(t,x) = \sin(nt)\sin(nx)/n^{p+1},
$$
$$
w(t,x) = \sinh(nt)\sin(nx)/n^{p+1},
$$
with $p\geq 1$, $n$, constants, defined on $t \geq 0$, $x\in [0,1]$. They
are solutions of the Cauchy problem for wave equation and the Laplace
equation, respectively, with precisely the same Cauchy data on $t=0$, that is
$$
v_{tt} - v_{xx} =0,\quad
v|_{t=0}= 0, \quad
v_t|_{t=0} = \sin(nx)/n^p,
$$
$$
w_{tt}+ w_{xx}=0, \quad
w|_{t=0}=0, \quad
w_t|_{t=0} = \sin(nx)/n^p.
$$
As $n\to \infty$, the Cauchy data converges to zero in
$C^{p-1}([0,1])$.  In this limit, the solution of the wave equation
converges to zero, while the solution of the Laplace equation
diverges. The concept of well posedness is introduced in order to
capture this behavior of the wave equation's solution under high
frequency perturbations on its Cauchy data.}.)

Let $B({\R}^n)$ be a Banach space with norm $\| ~ \|$, whose elements
are vector valued functions from $\R^n$ to $\R^m$.  {\em The Cauchy
  problem (\ref{pdoc}) is well posed in $B({\R}^n)$ if given initial
  data $f(x) \in B({\R}^n)$ there exists a solution $u(t,x)$ which is
  unique in $B({\R}^n)$ for each $t \in [0,T)$, for some $T > 0$; and
  given any number $\epsilon >0$ there exists $\delta >0$ such that,
  for every data $\tilde f(x)\in B(\R^n)$ satisfying $\|\tilde f - f\|
  < \delta$ there exists a unique solution $\tilde u(t,x)\in B(\R^n)$
  for $t\in [0,\tilde T)\times \R^n$ for some $\tilde T>0$, with
  $|\tilde T- T| <\epsilon$, and satisfying $\|\tilde u(t) - u(t)\|
  <\epsilon$, for all $t\in [0,\min(\tilde T,T))$.} This means that
the solution depends continuously on the data in the norm $\|~\|$.

Well posedness is essentially a statement about the behavior of the
solutions of a Cauchy problem under high frequency perturbations of
the initial data. Here is where pseudodifferential calculus is most
useful to study solutions of the Cauchy problem. The high frequency
part of the solution can be determined by studying the higher order
terms in the asymptotic expansion of symbols.

A wide class of operators with well posed Cauchy problem is called
strongly hyperbolic. {\em A first order pseudodifferential system
  (\ref{pdoc}) is strongly hyperbolic if $p \in\psicl^1$ and the
  principal symbol is symmetrizable.} This means that there exists a
positive definite, Hermitian operator $H(t,x,\w)$ homogeneous of
degree zero in $\w$, smooth in all its arguments for $\w\neq 0$, such
that
\[
(H\un p_1 + \un p_1^{*} H ) \in S^0,
\]
where $\un p_1^{*}$ is the adjoint of the principal symbol $\un p_1$.

The definition summarizes all the hypotheses on quasilinear systems
needed to prove well posedness. It is the definition given in Sec.
3.3.1 in \cite{Kreiss89} for linear variable coefficient systems, and
the so-called symmetrizable quasilinear systems given in Sec. 5.2 in
\cite{Taylor91}.

Consider first order differential systems of the form $\partial_t u =
A^i(t,x)\partial_iu + B(t,x)u$. The symbol is $\un p(t,x,\w) = i
A^j(t,x)\w_j + B(t,x)$, and the principal symbol is $\un p_1(t,x,\w) =
iA^j \w_j$. If the matrices $A^i$ are all symmetric, then the system
is called symmetric hyperbolic. The symmetrizer $H$ is the identity,
and $\un p_1 +\un p_1^{*} =0$. The wave equation on a fixed
background, written as a first order system is an example of a
symmetric hyperbolic system. Well posedness for symmetric hyperbolic
systems can be shown without pseudodifferential calculus. The basic
energy estimate can be obtained by integration by parts in space-time.

If the matrices $A^i$ are symmetrizable, then the differential system
is called strongly hyperbolic. The symmetrizer $H=H(t,x,\w)$ is
assumed to depend smoothly on $\w$. Every symmetric hyperbolic system
is strongly hyperbolic. pseudodifferential calculus must be used to
show well posedness for variable coefficient strongly hyperbolic
systems that are not symmetric hyperbolic \cite{Taylor81}. The
definition given two paragraphs above is more general because the
symbol does not need to be a polynomial in $\w$. The definition given
above includes first order pseudodifferential reductions of second
order differential systems. These type of reductions are performed
with operators like $\Lambda$, $\lambda$, or $\ell$, defined in the
Appendix \ref{e}.

In the particular case of constant coefficient systems there exists in
the literature a more general definition of strong hyperbolicity
\cite{Gustafsson95,Kreiss89}. The principal symbol $\PP$ must have
only imaginary eigenvalues, and a complete set of linearly independent
eigenvectors. The latter must be uniformly linear independent in
$\omega \neq 0$ over the whole integration region. Kreiss's matrix
theorem (see Sec. 2.3 in \cite{Kreiss89}) says that this definition is
equivalent to the existence of a symmetrizer $H$.  Nothing is known
about the smoothness of $H$ with respect to $t$, $x$, and $\omega$.
The existence of this symmetrizer is equivalent to well posedness for
constant coefficient systems. However, the proof of well posedness
for variable coefficient and quasilinear systems does require the
smoothness of the symmetrizer. There are examples showing that this
smoothness does not follow from the previous hypothesis on eigenvalues
and eigenvectors of $\PP$. Because it is not known what additional
hypothesis on the latter could imply this smoothness, one has to
include it into the definition of strong hyperbolicity for
nonconstant coefficient systems.

A more fragile notion of hyperbolicity is called weak hyperbolicity,
where the operator $\PP$ has imaginary eigenvalues, but nothing is
required of its eigenvectors. Quasilinear weakly hyperbolic systems
are not well posed. The following example gives an idea of the
problem. The $2\times 2$ system $\partial_t u = A \partial_xu$ with
$t,x\in\R$ and
\[
A =\left(
\begin{array}{cc}
1&1\\
0&1
\end{array}
\right),
\]
is weakly hyperbolic. Plane wave solutions of the form $u(t,x) = \hat
u(t) e^{i\w\cdot x}$ satisfy $|\hat u(t)| \leq |\hat u(0)| (1+|\w|t)$.
Therefore, plane wave solutions to a weakly hyperbolic system do not
diverge exponentially in the high frequency limit (as in the case of
Cauchy problem for the Laplace equation) but only polynomically. This
divergence causes solutions to variable coefficient weak hyperbolic
systems to be unstable under perturbations in the lower order terms of
the operator, as well as in the initial data.

The main theorem about well posedness for strongly hyperbolic
systems is the following. {\em The Cauchy problem (\ref{pdoc}) for a
  strongly hyperbolic system is well posed with respect to the Sobolev
  norm $\| ~ \|_s$ with $s > n/2+1$.  The solution belongs to
  $C([0,T),H^s)$, and $T>0$ depends only on $\|f\|_s$.}

In the case of strongly hyperbolic differential systems, this is
Theorem 5.2.D in \cite{Taylor91}. The proof for pseudodifferential
strongly hyperbolic systems is essentially the same. One builds an estimate
for the solution in a norm, defined using the symmetrizer, equivalent to
the Sobolev norm $H^s$. Then the argument follows the standard proof for
differential systems. The construction of the symmetrizer is basically the
one carried out in \cite{hKoOoR98}.

\section{ADM decomposition of Einstein's equation}
\label{S:GR}

The ADM decomposition of Einstein's equation is reviewed. The
densitized lapse function is introduced in Sec. \ref{s:ADM}. Theorem
\ref{T:ADMwh} says that the resulting evolution equations are weakly
hyperbolic, for every choice of a positive densitized lapse function
and spacelike shift vector as given fields. The BSSN-type system is
introduced in Sec. \ref{s:BSSN}.  It is essentially the system given
in \cite{mStN95,tBsS98} without conformal-traceless decomposition,
keeping the three connection functions and densitizing the lapse
function. It is reduced to a first order pseudodifferential system
like the ADM evolution equations. The main result, Theorem
\ref{T:BSSNsth}, asserts that BSSN-type equations are strongly
hyperbolic, for every positive densitized lapse function and spacelike
shift vector.

Let $(M,g_{ab})$ be a space-time solution of Einstein's equation.
That is a four-dimensional, smooth, orientable manifold $M$, and a
smooth, Lorentzian metric $g_{ab}$ solution of
\[
G_{ab}= \kappa T_{ab},
\]
with $G_{ab} = R_{ab} -Rg_{ab}/2$ the Einstein tensor, $T_{ab}$ the
stress-energy tensor, and $\kappa = 8\pi$. Ricci's tensor is $R_{ab}$
and $R$ denotes Ricci's scalar. Latin indices $a$, $b$, $c$, $d$
denote abstract indices, and they are raised and lowered with $g^{ab}$
and $g_{ab}$, respectively, with $g_{ac}g^{cb} = \delta_a{}^b$. The
unique torsion-free metric connection is denoted by $\nabla_a$. The
conventions throughout this work are $2\nabla_{[a}\nabla_{b]} v_c =
R_{abc}{}^dv_d$ for Riemann's tensor and $(-,+,+,+)$ for the metric
signature.

Prescribe on $M$ a foliation of spacelike hypersurfaces by introducing
a time function $t$, which is a scalar function satisfying the
condition that $\nabla_a t$ is everywhere timelike. Denote the
foliation by $S_t$, and by $n_a$ the unit normal to $S_t$ such that
$n^a = g^{ab}n_b$ is future directed. Therefore, $n_a = -N \nabla_at$
for some positive function $N$. Fixing the foliation determines its
first and second fundamental forms $h_{ab} = g_{ab} +n_a n_b$ and
$k_{ab} = - h_a{}^c \nabla_cn_b$, respectively. Decompose Einstein's
equation into evolution equations (\ref{eq:fforms}), (\ref{eq:ADM}),
and constraint equations (\ref{eq:MC}), (\ref{eq:HC}), as
follows\goodbreak
\begin{equation}
\label{eq:fforms}
\lie_n h_{ab} = -2 k_{ab},
\end{equation}
\begin{eqnarray}
\nonumber
\lie_n k_{ab} &=& {}^{(3)}R_{ab} - 2 k_a{}^ck_{bc} + k\,k_{ab} \\
\label{eq:ADM}
&& - (D_aD_b N)/N -\kappa S_{ab},
\end{eqnarray}
\begin{equation}\label{eq:MC}
D_b k_a{}^b-D_a k =  \kappa j_a,
\end{equation}
\begin{equation}\label{eq:HC}
{}^{(3)}R + k^2 - k_{ab} k^{ab} = 2\kappa \rho,
\end{equation}
where $\lie_n$ denotes the Lie derivative along $n^a$, and $k = k_a{}^a$.
Here ${}^{(3)}R_{ab}$, ${}^{(3)}R$, and $D_a$ are, respectively, the Ricci
tensor, the Ricci scalar, and the Levi-Civit\`a connection of $h_{ab}$,
while $h^{ab}$ denotes its inverse. The stress-energy tensor is
decomposed as $S_{ab} = \left( h_a{}^{c}h_b{}^{d} T_{cd} - T
h_{ab}/2\right)$, with $T= T_{ab} g^{ab}$, $j_a = -h_a{}^{c} n^{d}
T_{cd}$, and $\rho = T_{ab}n^a n^b$.

Introduce on $M$ a future-directed timelike vector field $t^a$.
Impose the additional condition $t^a \nabla_a t=1$, that is, the
integral lines of $t^a$ are parametrized precisely by $t$. This
condition implies that the orthogonal decomposition of $t^a$ with
respect to $S_t$ has the form $t^a = Nn^a + \beta^a$, with
$n_a\beta^a=0$. $N$ is called the lapse function and $\beta^a$ the
shift vector. The integral lines of $t^a$ determine a diffeomorphism
among the hypersurfaces $S_t$. This, in turn determines a coordinate
system on $M$ from a coordinate system on $S_0$.  Lie derivatives with
respect to $n^a$ can be rewritten in terms of $t^a$ and $\beta^a$. The
resulting equations are called the ADM decomposition of Einstein's
equation.

\subsection{Densitized ADM equations}
\label{s:ADM}

Consider the ADM decomposition of Einstein's equation.  Let $x^{\mu}$
be a coordinate system adapted to the foliation $S_t$, where $x^0 = t$
and $x^i$ are intrinsic coordinates on each $S_t$ that remain constant
along the integral lines of $t^a$.  Greek indices take values
$0,1,2,3$, and latin indices $i$, $j$, $k$, $l$, take values $1,2,3$.
In these coordinates, $t^{\mu} = \delta_0{}^{\mu}$, $n_{\mu} =-N
\delta_{\mu}{}^0$; then $\beta^{\mu} = \delta_i{}^{\mu} \beta^i$ and
$n^{\mu} = (\delta_0{}^{\mu} - \beta^{\mu})/N$. The components of the
space-time metric have the form
\[
g_{\mu\nu} = -N^2 \delta_{\mu}{}^0\delta_{\nu}{}^0 
+ h_{ij} (\beta^i \delta_{\mu}{}^{0}+\delta_{\mu}{}^{i})
(\beta^j \delta_{\nu}{}^{0}+\delta_{\nu}{}^{j}).
\]
In these coordinates
\begin{eqnarray*}
{}^{(3)}R_{ij} &=& \frac{1}{2} h^{kl} \left[
- \partial_k\partial_lh_{ij} -\partial_i\partial_j h_{kl}
+ 2 \partial_k\partial_{(i}h_{j)l} \right]  \\
&& + \gamma_{ikl}\gamma_j{}^{kl} - \gamma_{ij}{}^k\gamma_{kl}{}^l,
\end{eqnarray*}
where $\gamma_{\mu\nu}{}^{\sigma} = h_{\mu}{}^{\mu'} h_{\nu}{}^{\nu'}
h_{\sigma'}{}^{\sigma} \Gamma_{\mu'\nu}{}^{\sigma'}$ are the spatial
components of the Christoffel symbols of $g_{\mu\nu}$, and
$\gamma_i{}^{jk} := \gamma_{il}{}^kh^{jl}$.

Densitize the lapse function, that is, write it as $N = (\h)^b Q$,
where $\h := \sqrt{\det(h_{ij})}$, $b$ is a constant, and $Q$ is a
given, positive function. This modifies the principal part of Eq.
(\ref{eq:ADM}).  New terms containing second spatial derivatives of
$h_{ij}$ come from $(D_iD_jN)/N$.

Summarizing, the unknowns for the densitized ADM equations are
$h_{ij}$ and $k_{ij}$. The evolution equations,
Eqs. (\ref{eq:fforms}), (\ref{eq:ADM}), have the form\goodbreak
\begin{equation}\label{eq:ADMd1}
\lie_{(t-\beta)} h_{ij} = -2 N k_{ij},
\end{equation}
\begin{eqnarray}
\nonumber
\lie_{(t-\beta)} k_{ij} 
&=& (N/2) h^{kl} \left[ - \partial_k\partial_lh_{ij} 
- (1+b) \partial_i\partial_j h_{kl} \right. \\
\label{eq:ADMd2}
&& \left. + 2 \partial_k\partial_{(i}h_{j)l} \right] + B_{ij},
\end{eqnarray}
where $\lie_{(t-\beta)}h_{ij} = \partial_t h_{ij} - (\beta^k\partial_k
h_{ij} + 2 h_{k(i}\partial_{j)}\beta^k)$, and the same holds for
$k_{ij}$. The nonprincipal part terms are grouped in
\begin{eqnarray*}
B_{ij} &=& N \left[ \gamma_{ikl}\gamma_j{}^{kl} 
- \gamma_{ij}{}^k\gamma_{kl}{}^l - 2 k_i{}^lk_{jl} \right.\\
&& \left. + k_{ij} k_l{}^l -A_{ij}  -\kappa S_{ij} \right],\\
A_{ij} &=& a_ia_j - \gamma_{ij}{}^ka_k
+ \partial_i \partial_j (\ln Q) + 2b \gamma_{ikl}\gamma_j{}^{(kl)},
\end{eqnarray*}
with $a_{\mu} = n^{\nu}\nabla_{\nu} n_{\mu}= D_{\mu} (\ln N)$.
The relations $\ln N = b \ln \h +\ln Q$ and 
$(D_iD_jN)/N = (b/2) h^{kl} \partial_i \partial_j h_{kl} +A_{ij}$
were used.

The following result asserts that densitized ADM evolution equations
are weakly hyperbolic.
\begin{theorem}\label{T:ADMwh}
Fix any positive function $Q$, a vector field $\beta^i$,
and first and second fundamental forms $h_{ij}$, $k_{ij}$ 
on $S_0$. If $b \geq 0$, then 
Eqs. (\ref{eq:ADMd1}), (\ref{eq:ADMd2}) are weakly hyperbolic.
If $b < 0$, these equations are not hyperbolic.
\end{theorem}

The proof has two steps: first, to write down Eqs.  (\ref{eq:ADMd1}),
(\ref{eq:ADMd2}) as an appropriate first order pseudodifferential
system, Eqs. (\ref{eq:ADMpd1}), (\ref{eq:ADMpd2}); second, to split
the corresponding principal symbol into orthogonal parts with respect
to the Fourier variable $\omega_i$, and then to explicitly compute the
associated eigenvalues and eigenvectors.

{\bf Proof.} {\em First order reduction.}  Compute the symbol
associated with the second order operator given by Eqs.
(\ref{eq:ADMd1}), (\ref{eq:ADMd2}), that is,
\[
\partial_t h_{ij} 
= \int_{S_t} \{ -2N \hat k_{ij} + i \omega_k\beta^k \hat h_{ij}
+ 2 \hat h_{l(i} \partial_{j)}\beta^l\} e^{i\omega x} \dbar \omega,
\]
\begin{eqnarray*}
\partial_t k_{ij} &=&  \int_{S_t} \left\{ (N/2) \left[
|\omega|_h^2 \hat h_{ij} + (1+b)\omega_i\omega_j  h^{kl} \hat h_{kl}
\right. \right. \\
&&  \left. \left. - 2 \omega^k\omega_{(i}\hat h_{j)k} \right] 
+ i \omega_k\beta^k \hat k_{ij} + \tilde B_{ij} \right\} e^{i\omega x}
\dbar \omega,
\end{eqnarray*}
where $\hat h_{ij}$ and $\hat k_{ij}$ denote the Fourier transforms in
$x^i$ of $h_{ij}$ and $k_{ij}$, and 
\[
\tilde B_{ij} = \hat B_{ij} + 2 \hat k_{l(i}\partial_{j)}\beta^l
\]
denotes the terms not in the principal symbol. Here $\dbar \omega = d
\omega/(2\pi)^{3/2}$, $|\omega|_h^2 = \omega_i\omega_j h^{ij}$, and we
will use the convention $\omega^i = \omega_j h^{ij}$. Transform this
second order symbol into a first order one via $\hat \ell_{ij} = i
|\omega|_{\delta} \hat h_{ij}$, where $|\omega|_{\delta}^2 = \omega_i
\omega_j \delta^{ij}$, with $\delta^{ij} = \mbox{diag}(1,1,1)$. The
associated first order system is then
\begin{eqnarray}
\nonumber
\partial_t \ell_{ij} &=& \int_{S_t} \left\{
i |\omega|_h \left[ - (2N/\alpha) \hat k_{ij}
+ \tilde \beta \hat \ell_{ij}  \right] \right. \\
&& \label{eq:ADMpd1}
\left. + 2 \hat \ell_{k(i} \partial_{j)} \beta^k \right\}
e^{i\omega x}\dbar \omega,
\end{eqnarray}
\begin{eqnarray}
\nonumber
\partial_t k_{ij} &=& \int_{S_t} \left\{
i |\omega|_h \left[ -(N\alpha/2) \left( \hat \ell_{ij}
+(1+b)\tilde \omega_i\tilde \omega_j h^{kl}\hat \ell_{kl}
\right. \right. \right. \\
\label{eq:ADMpd2}
&& \left. \left. \left.
- 2 \tilde \omega^k \tilde \omega_{(i}\hat \ell_{j)k} \right)
+ \tilde \beta \hat k_{ij} \right]+ \tilde B_{ij} \right\}
e^{i\omega x} \dbar \omega,
\end{eqnarray}
with $\alpha = |\omega|_h/|\omega|_{\delta}$,
$\tilde \omega_i = \omega_i/|\omega|_h$,
$\tilde \beta:= \tilde \omega_k\beta^k$, and
$\ell_{ij} = \int_{S_t} i|\omega|_{\delta} \hat h_{ij}
e^{i\omega x} \dbar \omega$.
Then the symbol of equations above can be written as
\begin{equation}\label{eq:main}
p(t,x,u,i\omega) = i |\omega|_h \PP(t,x,u,\omega) 
+ \BB(t,x,u,\omega),
\end{equation}
where $(\BB \hat u)^T:= (2\hat \ell_{k(i} \partial_{j)} \beta^k,\tilde
B_{ij})$, $\hat u^T := (\hat \ell_{ij}, \hat k_{ij})$, with the upper
index $T$ meaning transpose. The principal part operator $\PP$ can be
read out from the terms inside the square brackets in Eqs.
(\ref{eq:ADMpd1}), (\ref{eq:ADMpd2}). Notice that the definition of
the principal symbol here differs from the one given in Sec.
\ref{S:wp} by a factor of $i|\omega|_h$.  (In particular, the
eigenvalues of $\PP$ as defined here must be real to be hyperbolic.)

{\em Eigenvalues and eigenvectors of $\PP$.} Once the principal
symbol is known, it only remains to compute its eigenvalues and
eigenvectors. The assumption $\alpha =1$ facilitates the computations.
It is not a restriction since the norms $|~|_{\delta}$ and $|~|_h$ are
equivalent and smoothly related, and therefore the properties of the
eigenvalues and eigenvectors of the principal symbol are the same with
either norm. Furthermore, one can check that if $\hat u^T = (\hat
\ell_{ij}, \hat k_{ij})$ is an eigenvector of $\PP (\alpha=1)$ with 
eigenvalue $\lambda$, then $\hat u^T (\alpha)= (\alpha^{-1/2} \hat
\ell_{ij}, \alpha^{1/2} \hat k_{ij})$ is an eigenvector of $\PP(\alpha)$ 
with the same eigenvalue $\lambda$.  Therefore, from now on $\alpha =1$ is
assumed. A second suggestion for doing these calculations is to
decompose the eigenvalue equation $\PP \hat u = \lambda \hat u$ into
orthogonal components with respect to $\tilde \omega_i$. Introduce the
splitting
\begin{equation}
\label{eq:splith}
\hat \ell_{ij} = \tilde \omega_i \tilde \omega_j \hat \ell
+ \hat \ell' q_{ij}/2
+ 2 \tilde \omega_{(i}\hat \ell'_{j)}
+ \hat \ell'_{\langle ij \rangle},
\end{equation}
\begin{equation}
\label{eq:splitk}
\hat k_{ij} = \tilde \omega_i \tilde \omega_j \hat k
+ \hat k^{\prime}q_{ij}/2
+ 2 \tilde \omega_{(i}\hat k_{j)}^{\prime}
+ \hat k^{\prime}_{\langle ij \rangle},
\end{equation}
where $q_{ij} := h_{ij} - \tilde \omega_i\tilde \omega_j$ is the
orthogonal projector to $\tilde \omega_i$, and
\[
\hat \ell = \tilde \omega^i \tilde \omega^j \hat \ell_{ij}, \quad
\hat \ell' =  q^{ij} \hat \ell_{ij}, \quad
\hat \ell'_i = q_i{}^k \tilde \omega^l \hat \ell_{kl},
\]
\[
\hat \ell'_{\langle ij \rangle} = q_i{}^k q_j{}^l \left( \hat \ell_{kl}
- \hat \ell' q_{kl}/2 \right).
\]
The same definitions hold for the $\hat k_{ij}$ components.
This decomposition implies that
$\hat u = \hat u^{(1)} + \hat u^{(2)} + \hat u^{(3)}$ where
\[
\hat u^{(1)} = \left[
\begin{array}{c}
\tilde \omega_i \tilde \omega_j \hat \ell + (q_{ij}/2) \hat \ell' \\
\tilde \omega_i \tilde \omega_j \hat k + (q_{ij}/2) \hat k'
\end{array} \right],
\]
\begin{equation}
\label{decomp1}
\hat u^{(2)} = \left[
\begin{array}{c}
2 \tilde \omega_{(i} \hat \ell'{}_{j)} \\
2 \tilde \omega_{(i} \hat k'{}_{j)} 
\end{array} \right],\quad
\hat u^{(3)} = \left[
\begin{array}{c}
\hat \ell'_{\langle ij\rangle} \\
\hat k'_{\langle ij \rangle} 
\end{array} \right].
\end{equation}
The principal symbol $\PP$ and the eigenvalue equation $\PP \hat u =
\lambda \hat u$ can also be decomposed into the same three parts. The
first part is four dimensional, corresponding to the variable $\hat
u^{(1)}$, that is, the scalar fields, $\hat \ell$, $\hat k$, $\hat
\ell'$, and $\hat k'$. The eigenvalues are
\[
\tilde \lambda^{(1)}_1 = \pm 1, \quad \tilde \lambda^{(1)}_2 = \pm \sqrt{b},
\]
where $\tilde \lambda := (\lambda-\tilde \beta)/N$, so the role of the
shift vector is to displace the value of the eigenvalue by an amount
$\tilde \beta = \tilde \omega_k\beta^k$, and the lapse rescales it.
But a change of lapse (which here is the function $Q$) and shift
cannot change a real eigenvalue into an imaginary one. It cannot
affect the hyperbolicity of the system. The associated eigenvectors
for this first part are
\[
\hat u^{(1)}_{\lambda_1} = \left[
\begin{array}{c}
2[(1+b) \tilde \omega_i \tilde \omega_j + (1-b) q_{ij}/2] \\
\mp [(1+b) \tilde \omega_i \tilde \omega_j + (1-b) q_{ij}/2] 
\end{array}\right],
\]
\[
\hat u^{(1)}_{\lambda_2} = \left[
\begin{array}{c}
2 \tilde \omega_i \tilde \omega_j \\
\mp \sqrt{b} \, \tilde \omega_i \tilde \omega_j
\end{array}\right].
\]
Notice that for $b=1$ the two eigenvectors $\hat u^{(1)}_{\lambda_1}$
collapse to the two eigenvectors $\hat u^{(1)}_{\lambda_2}$.  The
conclusion for this part is that the eigenvalues are real for $b\geq
0$, and the four eigenvectors are linearly independent for $b\neq 0$,
$b\neq 1$.

The second part is also four dimensional and corresponds to the
variable $\hat u^{(2)}$, that is, the vector fields $\hat \ell'_i$ and
$\hat k'_i$.  (The vector $\hat \ell'_i$ has only two independent
components because of the condition $\hat \ell'_i\tilde \omega^i=0$.
The same holds for $\hat k'_i$.) The result is
\[
\tilde \lambda^{(2)}_1 =  0,\quad
\hat u^{(2)}_{\lambda_1} = \left[
\begin{array}{c}
v_{j}{}^A \\
0
\end{array}\right].
\]
The eigenvalue has multiplicity 4, but there are only two linearly
independent eigenvectors. Here, $v_{j}{}^A$ represent two linearly
independent vectors, each one orthogonal to $\tilde \omega_i$, and
labeled with the index $A$, which takes values $1,2$. This part is the
main reason why the ADM equations are weakly hyperbolic.

The last part is again four dimensional and corresponds to the
variable $\hat u^{(3)}$, that is, the two-tensor fields $\hat
\ell'_{\langle ij \rangle}$ and $\hat k'_{\langle ij \rangle}$.  (The
tensor $\hat \ell'_{\langle ij \rangle}$ has only two independent
components because of the symmetry, the orthogonality to $\tilde
\omega_i$, and the trace-free condition. The same holds for $\hat
k'_{\langle ij \rangle}$.)  The result is
\[
\tilde \lambda^{(3)}_1 =  \pm 1, \quad
\hat u^{(3)}_{\lambda_1} = \left[
\begin{array}{c}
2 v_{\langle ij\rangle}{}^A \\
\mp v_{\langle ij \rangle}{}^A 
\end{array} \right].
\]
The eigenvalues each have multiplicity 2, and there are four linearly
independent eigenvectors. Here $v_{\langle kl \rangle}{}^A$ represent
two linearly independent symmetric, traceless tensors, orthogonal to
$\tilde \omega_i$.

At the end one gets the following picture. All eigenvalues are real
for $b\geq 0$. Notice that $\tilde \lambda^{(1)}_2$ becomes imaginary
for $b <0$, so the equations are not hyperbolic in this case. With
respect to the eigenvectors, there are two main cases.  First, $b> 0$
and $b\neq 1$. Then, the eigenvectors of the first and third parts of
$\PP$ do span their associated eigenspaces; but the eigenvectors $\hat
u^{(2)}_{\lambda}$ corresponding to the second part of $\PP$ do not
span their eigenspace. In the second case, $b=0$ or $b=1$. In this
case there are linearly dependent eigenvectors even among the scalar
variables. Therefore, the conclusion is that the system
(\ref{eq:ADMd1}), (\ref{eq:ADMd2}) is weakly hyperbolic for $b\geq
0$.\qed

It is interesting here to comment on the role of the Hamiltonian
constraint. Suppose that a term of the form $a h_{ij}$ times Eq.
(\ref{eq:HC}) is added to Eq. (\ref{eq:ADM}). Here $a$ is some real
constant. Can this modification alter the hyperbolicity of the ADM
equations? One might think that adding the Hamiltonian constraint to
the ADM evolution equation could have a similar role as densitizing
the lapse function i.e., it could keep both eigenvectors $\hat
u^{(1)}_{\lambda_2}$ linearly independent. The fact is, it does not.
Such an addition of the Hamiltonian constraint modifies only $\tilde
\lambda^{(1)}_1$ and $\hat u^{(1)}_{\lambda_1}$, and does not modify
$\hat u^{(1)}_{\lambda_2}$ and $\tilde \lambda^{(1)}_2$. The result is
\[
\tilde \lambda^{(1)}_1 = \pm \sqrt{1+2a},
\]
\[
\hat u^{(1)}_{\lambda_1} = \left[
\begin{array}{c}
2[(1+b+2a) \tilde \omega_i \tilde \omega_j + (1-b+2a) q_{ij}/2] \\
\mp \sqrt{1+2a}\,[(1+b) \tilde \omega_i \tilde \omega_j + (1-b) q_{ij}/2] 
\end{array}\right].
\]
Therefore, adding the Hamiltonian constraint only helps to keep the
eigenvectors $\hat u^{(1)}_{\lambda_1}$ independent of the $\hat
u^{(1)}_{\lambda_2}$, so it helps only in the case $b=1$, where the
former collapse onto the latter (for $a=0$). For $b\neq 1$ the
addition of the Hamiltonian constraint does not contribute to make the
vectors $\hat u^{(1)}_{\lambda_2}$ linearly independent, whereas
densitizing the lapse does.

\subsection{BSSN-type equations}
\label{s:BSSN}

Consider the densitized ADM evolution equations
(\ref{eq:ADMd1}), (\ref{eq:ADMd2}). Introduce into these equations
the new variable
\[
f^{\mu} := h^{\nu\sigma}\gamma_{\nu\sigma}{}^{\mu}.
\]
By definition $n_{\mu}f^{\mu} =0$, that is, $f^0=0$, so the new
variables are the components $f^i = h^{ij}[h^{kl} \partial_k h_{lj} -
\partial_j (\ln \h)]$, where $\h = \sqrt{\det (h_{ij})}$ as above.
They are related to the three connection variables $\tilde \Gamma^i$
of the BSSN system defined in Eq. (21) in \cite{tBsS98}. More
precisely, $\tilde \gamma_{ij}\tilde \Gamma^j = f_i +
(1/3)\partial_i(\ln \h)$, where $\tilde \gamma_{ij}$ is defined in Eq.
(10) of that reference.  The evolution equation for $f_i$ is obtained
by taking the trace in indices $\nu$, $\sigma$ of the identity
\begin{eqnarray*}
h_{\mu\delta} \lie_n \gamma_{\nu\sigma}{}^{\delta}
&=& -2 D_{(\nu}k_{\sigma)\mu} + D_{\mu} k_{\nu\sigma}
- 2 a_{(\nu}k_{\sigma)\mu} \\
&& + k_{\nu\sigma}a_{\mu}
+ \frac{1}{N} h_{\mu\delta} h_{\nu}{}^{\nu'} h_{\sigma}{}^{\sigma'}
\partial_{\nu'}\partial_{\sigma'} \beta^{\delta},
\end{eqnarray*}
where $\lie_n \gamma_{\nu\sigma}{}^{\delta} = n^{\mu} \partial_{\mu}
\gamma_{\nu\sigma}{}^{\delta} + 2 \gamma_{\mu(\nu}{}^{\delta}
\partial_{\sigma)}n^{\mu} - \gamma_{\nu\sigma}{}^{\mu}
\partial_{\mu}n^{\delta}$, and adding to the result $c$ times the
momentum constraint (\ref{eq:MC}). Here $c$ is any real constant. One
then gets
\[
\lie_n f_{\mu} = (c-2) D_{\nu}k_{\mu}{}^{\nu} + (1- c) D_{\mu} k + C_{\mu},
\]
where the nonprincipal terms are grouped in
\begin{eqnarray*}
C_{\mu} &=& -c \kappa j_{\mu} - 2 k_{\mu\nu}a^{\nu}
+ ka_{\mu} -2 \gamma_{\nu\sigma\mu}k^{\nu\sigma} \\
&& - 2 k_{\mu\nu}f^{\nu} 
+ (1/N) h_{\mu\nu} h^{\sigma\delta} \partial_{\sigma}\partial_{\delta}
\beta^{\nu}.
\end{eqnarray*}

Summarizing, the unknowns for BSSN-type systems are $h_{ij}$, $k_{ij}$,
and $f_{i}$. The evolution equations are
\begin{equation}\label{eq:SNd1}
\lie_{(t-\beta)} h_{ij} = -2 N k_{ij},
\end{equation}
\begin{eqnarray}
\nonumber
\lie_{(t-\beta)} k_{ij} &=& \frac{N}{2} h^{kl}
\left[ -\partial_k \partial_l h_{ij}
- b \, \partial_i\partial_j h_{kl} \right] \\
\label{eq:SNd2}
&& + N \partial_{(i}f_{j)} + \B_{ij},
\end{eqnarray}
\begin{equation}\label{eq:SNd3}
\lie_{(t-\beta)}f_i = N [ (c-2) h^{kj} \partial_k k_{ij} 
+(1-c) h^{kj} \partial_i k_{kj} ] + \C_i,
\end{equation}
where $\lie_{(t-\beta)} f_i = \partial_t f_i -(\beta^j\partial_j f_i
+ f_j \partial_i \beta^j)$, while $\lie_{(t-\beta)}h_{ij}$ and
$\lie_{(t-\beta)} k_{ij}$ are defined below
Eqs.(\ref{eq:ADMd1})-(\ref{eq:ADMd2}), and
\begin{eqnarray*}
\B_{ij} &=& N \left[ 
2 \gamma_{kl(i} \gamma_{j)}{}^{kl} + \gamma_{ikl}\gamma_j{}^{kl}  
- \gamma_{ijl} \gamma_{k}{}^{kl}  \right. \\
&& \left.
- 2 k_i{}^lk_{jl} + k_{ij} k_l{}^l -A_{ij}  -\kappa S_{ij} \right], \\
\C_i &=& N [C_i + (c-2) (\gamma_{kj}{}^k k_i{}^j
- \gamma_{ki}{}^j k_j{}^k)].
\end{eqnarray*}
The constraint equations are Eqs. (\ref{eq:MC}), (\ref{eq:HC}) and
\[
f^{\mu} - h^{\nu\sigma} \gamma_{\nu\sigma}{}^{\mu} =0.
\]

The main result of this work asserts that BSSN-type evolution equations
are strongly hyperbolic for some choices of the free parameters. 
\begin{theorem} \label{T:BSSNsth}
Fix any positive function $Q$, vector field $\beta^i$,
first and second fundamental forms $h_{ij}$, $k_{ij}$ on $S_0$.

If $b > 0$, $b\neq 1$, and $ c> 0$, then
Eqs. (\ref{eq:SNd1})-(\ref{eq:SNd3})  are strongly hyperbolic.

Assume that $b=1$. If $c=2$, then Eqs.
(\ref{eq:SNd1})-(\ref{eq:SNd3}) are strongly hyperbolic;
if $c \neq 2$, $c >0$, then they are weakly hyperbolic.
\end{theorem}

One can check that the system (\ref{eq:SNd1})-(\ref{eq:SNd3}) remains
strongly hyperbolic under a transformation of the form $F_i = f_i + d
\,\partial_i(\ln \h)$ for any real constant $d$, in particular
$d=1/3$, which gives the BSSN variable $\tilde \Gamma^i$.  (See the
comment at the end of this section.)

The first part of the proof follows the argument that establishes Theorem
\ref{T:ADMwh}. That is, from Eqs. (\ref{eq:SNd1})-(\ref{eq:SNd3}) obtain the
first order pseu\-do\-dif\-fer\-en\-tial system Eqs.
(\ref{eq:pseudo1})-(\ref{eq:pseudo3}) below. Then compute the eigenvector
and eigenvalues, by splitting the principal symbol into orthogonal parts
with respect to the Fourier variable $\omega_i$. Finally, the second part of
the proof is the construction of the symmetrizer.

{\bf Proof.} {\em First order reduction.}
Compute the symbol associated with the second order
operator given by Eqs. (\ref{eq:SNd1})-(\ref{eq:SNd3}):
\[
\partial_t h_{ij} = \int_{S_t} \left\{ -2 N \hat k_{ij} 
+ i  \omega_k\beta^k \hat h_{ij} 
+ 2 \hat h_{k(i} \partial_{j)} \beta^k \right\} e^{i\omega x} \dbar \omega,
\]
\begin{eqnarray*}
\partial_t k_{ij} &=& \int_{S_t} \left\{ (N/2) \left[
|\omega|_h^2 \hat h_{ij} + b \, \omega_i\omega_j h^{kl} \hat h_{kl}
\right] \right. \\
&& \left. + i N \omega_{(i}\hat f_{j)}
+ i \omega_k\beta^k \hat k_{ij}+ \tilde \B_{ij} \right\}
e^{i\omega x} \dbar \omega,
\end{eqnarray*}
\begin{eqnarray*}
\partial_t f_i &=& \int_{S_t} \left\{
iN \left[ (c-2) \hat k_{ik} \omega^k
+(1-c) \omega_i h^{kj} \hat k_{kj} \right] \right. \\
&& \left. + i \omega_k \beta^k \hat f_i + \tilde \C_i\right\}
e^{i\omega x} \dbar \omega,
\end{eqnarray*}
where the terms not in the principal symbol have the forms
\[
\tilde \B_{ij} = \hat \B_{ij} + 2 \hat k_{k(i} \partial_{j)} \beta^k,
\]
\[
\tilde \C_i = \hat \C_i + \hat f_k \partial_i \beta^k.
\]
Here $\dbar \omega = d \omega/(2\pi)^{3/2}$, $|\omega|_h^2 =
\omega_i\omega_j h^{ij}$, and $\omega^i = \omega_j h^{ij}$.  Introduce
the unknown $\hat \ell_{ij} = i |\omega|_{\delta} \hat h_{ij}$, with
$|\omega|_{\delta}^2 = \omega_i \omega_j \delta^{ij}$, where
$\delta^{ij} = \mbox{diag}(1,1,1)$. The resulting pseudodifferential
system is a first order one, given by
\begin{eqnarray}
\nonumber
\partial_t \ell_{ij} &=& \int_{S_t} \left\{ i|\omega|_h
\left[ -(2N/\alpha) \hat k_{ij} 
+ \tilde \beta \hat \ell_{ij} \right] \right. \\
\label{eq:pseudo1}
&&\left. + 2 \hat \ell_{k(i} \partial_{j)} \beta^k \right\}
e^{i\omega x} \dbar \omega,
\end{eqnarray}
\begin{eqnarray}
\nonumber
\partial_t k_{ij} &=& \int_{S_t} \left\{ i |\omega|_h \left[ (N\alpha/2)
\left( - \hat \ell_{ij}
- b\, \tilde \omega_i\tilde \omega_j h^{kl}\hat \ell_{kl}
\right.\right. \right.\\
\label{eq:pseudo2}
&& \left. \left. \left. + 2 \tilde \omega_{(i}\hat f_{j)} \right)
+ \tilde \beta \hat k_{ij} \right]
+ \tilde \B_{ij} \right\} e^{i\omega x} \dbar \omega,
\end{eqnarray}
\begin{eqnarray}
\nonumber
\partial_t f_{i} &=& \int_{S_t} \left\{ i |\omega|_h
\left[ N \left( (c-2) \hat k_{ik}\tilde \omega^k
\right. \right. \right.  \\
\label{eq:pseudo3}
&& \left. \left. \left. 
+(1-c) \tilde \omega_i h^{kj} \hat k_{kj} \right)
+ \tilde \beta \hat f_{i} \right]
+ \tilde \C_{i} \right\} e^{i\omega x} \dbar \omega,
\end{eqnarray}
with $\alpha = |\omega|_h / |\omega|_{\delta}$,
$\tilde \omega_i = \omega_i/|\omega|$,
$\tilde \beta = \tilde \omega_k \beta^k$, and
$\ell_{ij} = \int_{S_t} i |\omega|_{\delta} \hat h_{ij} e^{i \omega x}
\dbar \omega$.
The symbol of Eqs. (\ref{eq:pseudo1})-(\ref{eq:pseudo3}) has the form,
\begin{equation}\label{eq:pseudo-main}
p(t,x,u,i\omega) = i |\omega|_h \PP(t,x,u,\omega)
+ \BB(t,x,u,\omega),
\end{equation}
where $(\BB \hat u)^T :=(2\hat\ell_{k(i} \partial_{j)} \beta^k, \tilde
\B_{ij}, \tilde \C_{i})$, $\hat u^T := (\hat \ell_{ij},\hat k_{ij},
\hat f_{i})$, the index $T$ denotes the transpose, and the principal
symbol $\PP$ can be read out from the terms inside the square brackets
in Eqs. (\ref{eq:pseudo1})-(\ref{eq:pseudo3}). As in the proof of
Theorem \ref{T:ADMwh}, the definition of the principal symbol here
differs from the one given in Sec. \ref{S:wp} by a factor of
$i|\omega|_h$.

{\em Eigenvalues and eigenvectors of $\PP$.}  Following the proof of
Theorem \ref{T:ADMwh}, $\alpha =1$ is assumed.  One can check in this
case that, if $\hat u^T = (\hat \ell_{ij}, \hat k_{ij}, \hat f_i)$ is
an eigenvector of $\PP (\alpha=1)$ with eigenvalue $\lambda$, then
$\hat u^T (\alpha)= (\alpha^{-2/3} \hat \ell_{ij}, \alpha^{1/3} \hat
k_{ij}, \alpha^{1/3} \hat f_i)$ is an eigenvector of $\PP(\alpha)$
with the same eigenvalue $\lambda$.

The orthogonal decomposition (\ref{eq:splith}), (\ref{eq:splitk})
simplifies the calculation. In addition, decompose
\[
\hat f_i = \tilde \omega_i \hat f + \hat f'_i,
\]
where $\hat f = \tilde \omega_i \hat f^i$ and $\hat f'_i = q_i{}^j
\hat f_j$.  This decomposition implies that $\hat u = \hat u^{(1)} +
\hat u^{(2)} + \hat u^{(3)}$ where
\[
\hat u^{(1)} = \left[
\begin{array}{c}
\tilde \omega_i \tilde \omega_j \hat \ell + (q_{ij}/2) \hat \ell' \\
\tilde \omega_i \tilde \omega_j \hat k + (q_{ij}/2) \hat k' \\
\tilde \omega_i \hat f
\end{array} \right],
\]
\begin{equation}
\label{decomp}
\hat u^{(2)} = \left[
\begin{array}{c}
2 \tilde \omega_{(i} \hat \ell'{}_{j)} \\
2 \tilde \omega_{(i} \hat k'{}_{j)} \\
\hat f'_i
\end{array} \right], \quad
\hat u^{(3)} = \left[
\begin{array}{c}
\hat \ell'_{\langle ij\rangle} \\
\hat k'_{\langle ij \rangle} \\
0
\end{array} \right].
\end{equation}
Split the principal symbol $\PP$ and the eigenvalue equation $\PP \hat
u = \lambda \hat u$ into the same three parts. The first part is five
dimensional, corresponding to the variable $\hat u^{(1)}$, that is, the
scalar fields $\hat f$, $\hat \ell$, $\hat k$, $\hat \ell'$, and
$\hat k'$.  The eigenvalues are
\[
\tilde \lambda^{(1)}_1 = \pm 1, \quad \tilde \lambda^{(1)}_2 = \pm \sqrt{b},
\quad \tilde \lambda^{(1)}_3 = 0,
\]
each having multiplicity 1, where again $\tilde \lambda :=
(\lambda-\tilde \beta)/N$. Hence, the eigenvalues are real if $b\geq
0$. The corresponding eigenvectors are
\[
\hat u^{(1)}_{\lambda_1} = \left[
\begin{array}{c}
2[(b-c+1)\tilde \omega_i \tilde \omega_j + (1-b)(q_{ij}/2) ]\\
\mp[(b-c+1)\tilde \omega_i \tilde \omega_j + (1-b) (q_{ij}/2)] \\
(2-c) b \tilde \omega_i 
\end{array} \right],
\]
\[
\hat u^{(1)}_{\lambda_2} = \left[
\begin{array}{c}
2 \tilde \omega_i \tilde \omega_j \\
\mp \sqrt{b} \,\tilde \omega_i \tilde \omega_j \\
\tilde \omega_i 
\end{array} \right], \quad
\hat u^{(1)}_{\lambda_3} = \left[
\begin{array}{c}
2 \tilde \omega_i \tilde \omega_j \\
0 \\
(1+b) \tilde \omega_i
\end{array} \right].
\]
Notice that both eigenvectors $\hat u^{(1)}_{\lambda_2}$ collapse if
$b=0$.  Also see that the eigenvectors $\hat u^{(1)}_{\lambda_1}$
collapse to $\hat u^{(1)}_{\lambda_2}$ in the case $b=1$ and $c\ne 2$.
Thus, in these cases Eqs. (\ref{eq:pseudo1})-(\ref{eq:pseudo3}) are
weakly hyperbolic.  In the case $b=1$ and $c=2$ the eigenvalues $\pm
\sqrt{b}$ collapse to $\pm 1$. Therefore, one has $\tilde
\lambda^{(1)}_1 = \pm 1$, each with multiplicity 2, and $\tilde
\lambda^{(1)}_2 =0$, with multiplicity 1. There are five linearly
independent eigenvectors in this case,
\[
\hat u^{(1)}_{\lambda_1} = \left[
\begin{array}{c}
2 \tilde \omega_i \tilde \omega_j \\
\mp \tilde \omega_i \tilde \omega_j \\
\tilde \omega_i 
\end{array} \right], \quad
\hat u^{(1)}_{\lambda_1} = \left[
\begin{array}{c}
q_{ij} \\
\mp q_{ij}/2 \\
\tilde \omega_i 
\end{array} \right],
\]
\[
\hat u^{(1)}_{\lambda_2} = \left[
\begin{array}{c}
\tilde \omega_i \tilde \omega_j \\
0 \\
\tilde \omega_i
\end{array} \right].
\]

The second part is six dimensional and corresponds to the variables
$\hat u^{(2)}$, that is, the vector fields $\hat f'_i$, $\hat \ell'_i$,
and $\hat k'_i$, orthogonal to $\tilde \omega_i$. The eigenvalues are
\[
\tilde \lambda^{(2)}_1 =  0,\quad
\tilde \lambda^{(2)}_2 = \pm \sqrt{c/2},
\]
where each one has multiplicity 2. They are real if $c\geq 0$. There are
six linearly independent eigenvectors in the case $c > 0$, given by
\[
\hat u^{(2)}_{\lambda_1} = \left[
\begin{array}{c}
2 \tilde \omega_{(i} v_{j)}{}^A \\
0 \\
v_i{}^A
\end{array} \right], \quad
\hat u^{(2)}_{\lambda_2} = \left[
\begin{array}{c}
4 \tilde \omega_{(i} v_{j)}{}^A \\
\mp \sqrt{2c} \, \tilde \omega_{(i} v_{j)}{}^A \\
(2-c) v_i{}^A
\end{array} \right],
\]
where $v_j{}^A$ represent two linearly independent vectors, each one
orthogonal to $\tilde \omega_i$, labeled by the index $A$ which takes
values $1,2$. Here is the key role of the momentum constraint.  If
$c=0$, that is, the momentum constraint is not added to the system,
then the two eigenvectors $\hat u^{(2)}_{\lambda_2}$ become linearly
dependent, as occurs in the densitized ADM evolution equations.

The last part is four dimensional and is the same as in
Theorem \ref{T:ADMwh}. It corresponds to the variables $\hat u^{(3)}$,
that is, the tensor fields $\hat \ell'_{\langle ij \rangle}$ and
$\hat k'_{\langle ij \rangle}$. The result is
\[
\tilde \lambda^{(3)}_1 =  \pm 1,\quad
\hat u^{(3)}_{\lambda_1} = \left[
\begin{array}{c}
2 v_{\langle ij\rangle}{}^A \\
\mp v_{\langle ij \rangle}{}^A \\
0
\end{array} \right].
\]
The eigenvalues each have multiplicity 2, and there are four linearly
independent eigenvectors. The tensors $v_{\langle kl \rangle}{}^A$
represent two linearly independent symmetric, traceless tensors,
orthogonals to $\tilde \omega_i$.

{\em The symmetrizer.} The operator $H=(T^{-1})^{*} (T^{-1})$ is a
symmetrizer of system (\ref{eq:pseudo-main}), where $T$ is an operator
whose columns correspond to the eigenvectors of $\PP$, $T^{-1}$ is its
inverse, and $*$ denotes the adjoint. Then it only remains to do a
lengthy, although straightforward, calculation.

There are two hints that help to simplify the construction of the
symmetrizer. They are based on the observation that the principal
symbol has the particular form $\PP = N \tilde \PP + \tilde \beta
\Id$, where $\Id$ is the identity matrix $\Id = \mbox{diag}(h_{(i}{}^k
h_{j)}{}^l, h_{(i}{}^k h_{j)}{}^l, h_i{}^k)$, and
\[
\tilde \PP = \left[
\begin{array}{ccc}
0 & \tilde \PP_{(\ell)ij}{}^{(k)kl} & 0 \\
\tilde \PP_{(k)ij}{}^{(\ell)kl} & 0 & \tilde \PP_{(k)ij}{}^{(f)k} \\
0 & \tilde \PP_{(f)i}{}^{(k)kl} & 0
\end{array}
\right],
\]
where the indices $(\ell)$, $(k)$, and $(f)$ indicate rows and columns,
that is, equations and variables, respectively, and the matrix components
are given by
\begin{eqnarray*}
\tilde \PP_{(\ell)ij}{}^{(k)kl} &=& -2  h_{(i}{}^k h_{j)}{}^l,\\
\tilde \PP_{(k)ij}{}^{(\ell)kl} &=& (-1/2)[h_{(i}{}^k h_{j)}{}^l
+ b \, \tilde \omega_i \tilde \omega_j h^{kl} ],\\
\tilde \PP_{(k)ij}{}^{(f)k} &=&  \tilde \omega_{(i} h_{j)}{}^k, \\
\tilde \PP_{(f)i}{}^{(k)kl} &=&  (c-2) \tilde \omega^{(k}h^{l)}{}_i
+(1- c) \tilde \omega_i h^{kl}.
\end{eqnarray*}
Then, the first hint is that a symmetrizer for $\PP$ is indeed a symmetrizer
for the nondiagonal elements $\tilde \PP$. A second hint is that the
orthogonal decomposition in Eq. (\ref{decomp}) induces the same splitting in
$\tilde \PP = \tilde \PP^{(1)} + \tilde \PP^{(2)} + \tilde \PP^{(3)}$
and therefore in $H = H^{(1)} + H^{(2)} + H^{(3)}$.

The result is
\[
H = \left[
\begin{array}{ccc}
H_{(\ell)ij}{}^{(\ell)kl} & 0 & H_{(\ell)ij}{}^{(f)k} \\
0 & H_{(k)ij}{}^{(k)kl} & 0 \\
H_{(f)i}{}^{(\ell)kl} & 0 & H_{(f)i}{}^{(f)k} 
\end{array}\right],
\]
where\goodbreak
\begin{eqnarray*}
H_{(\ell)ij}{}^{(\ell)kl} &=&
H^{(1)}_{(\ell)ij}{}^{(\ell)kl} + H^{(2)}_{(\ell)ij}{}^{(\ell)kl}
+ H^{(3)}_{(\ell)ij}{}^{(\ell)kl},\\
H_{(\ell)ij}{}^{(f)k} &=&
H^{(1)}_{(\ell)ij}{}^{(f)k} + H^{(2)}_{(\ell)ij}{}^{(f)k}, \\
H_{(k)ij}{}^{(k)kl} &=&
H^{(1)}_{(k)ij}{}^{(k)kl} + H^{(2)}_{(k)ij}{}^{(k)kl}
+ H^{(3)}_{(k)ij}{}^{(k)kl}, \\
H_{(f)i}{}^{(\ell)kl} &=&
H^{(1)}_{(f)i}{}^{(\ell)kl} + H^{(2)}_{(f)i}{}^{(\ell)kl},\\
H_{(f)i}{}^{(f)k} &=& H^{(1)}_{(f)i}{}^{(f)k} + H^{(2)}_{(f)i}{}^{(f)k}.
\end{eqnarray*}
The matrix coefficients of each part depend on $\tilde \omega_i$.
The scalar variable part of the symmetrizer has the form
\begin{eqnarray*}
H^{(1)}_{(\ell)ij}{}^{(\ell)kl} &=&
H_{\ell\ell} \tilde \omega_i \tilde \omega_j \tilde \omega^k \tilde \omega^l
+ H_{\ell'\ell'} \frac{q_{ij}}{2} q^{kl} \\
&& + H_{\ell\ell'} \left(\tilde \omega_i \tilde \omega_j q^{kl}
+\frac{q_{ij}}{2} \tilde \omega^k \tilde \omega^l \right), \\
H^{(1)}_{(\ell)ij}{}^{(f)k} &=&
H_{f\ell} \tilde \omega_i \tilde \omega_j \tilde \omega^k
+ H_{f\ell'} \frac{q_{ij}}{2} \tilde \omega^k, \\
H^{(1)}_{(k)ij}{}^{(k)kl} &=&
H_{kk} \tilde \omega_i \tilde \omega_j \tilde \omega^k \tilde \omega^l
+ H_{k'k'} \frac{q_{ij}}{2} q^{kl} \\
&& + H_{kk'} \left(\tilde \omega_i \tilde \omega_j q^{kl}
+\frac{q_{ij}}{2} \tilde \omega^k \tilde \omega^l \right), \\
H^{(1)}_{(f)i}{}^{(\ell)kl} &=&
H_{f\ell} \tilde \omega_i \tilde \omega^k \tilde \omega^l
+ H_{f\ell'} \tilde \omega_i q^{kl}, \\
H^{(1)}_{(f)i}{}^{(f)k} &=& H_{ff} \tilde \omega_i \tilde \omega^k,
\end{eqnarray*}
where the scalar functions that appear above are the following:
\[
H_{\ell\ell} = \frac{1}{8b^2} [2+ (1+b)^2], \quad
H_{kk} = \frac{1}{2b},
\]
\[
H_{ff} = \frac{3}{2b^2},\quad
H_{k'k'} = \frac{(b+1-c)^2 + b}{2b(1-b)^2},
\]
\[
H_{\ell'\ell'} = \frac{1}{4b^2}
\left[ (c-1)^2 +\frac{(b^2+1-c)^2+b^2}{2(1-b)^2}\right], 
\]
\[
H_{\ell\ell'} = \frac{1}{4b^2}
\left[c-1 - \frac{(b^2+1-c)(1+b)}{2(1-b)}\right],
\]
\[
H_{f \ell'} = -\frac{1}{4b^2}\left[2(c-1)
- \frac{(b^2+1-c)}{(1-b)}\right],
\]
\[
H_{kk'} = \frac{(c-1-b)}{2b(1-b)},\quad
H_{f \ell} = -\frac{1}{4b^2}(b+3).
\]

The symmetrizer for the vector variable part is
\begin{eqnarray*}
H^{(2)}_{(\ell)ij}{}^{(\ell)kl} &=&
\frac{2 (c-2)^2 +1}{4c^2}
2 \tilde \omega_{(i}q_{j)}{}^{(k} \tilde \omega^{l)},\\
H^{(2)}_{(\ell)ij}{}^{(f)k} &=&
\frac{4 (c -2) -1}{2 c^2} \tilde \omega_{(i}q_{j)}{}^k,\\
H^{(2)}_{(k)ij}{}^{(k)kl} &=&
\frac{1}{2c} 2 \tilde \omega_{(i}q_{j)}{}^{(k} \tilde \omega^{l)},\\
H^{(2)}_{(f)i}{}^{(\ell)kl} &=&
\frac{4(c -2)-1}{2 c^2} \tilde \omega^{(k}q^{l)}{}_i,\\
H^{(2)}_{(f)i}{}^{(f)k} &=& \frac{9}{2c^2} q_i{}^k.
\end{eqnarray*}

Finally, the two-tensor part of the symmetrizer is 
\begin{eqnarray*}
H^{(3)}_{(\ell)ij}{}^{(\ell)kl} &=&
\frac{1}{8} q_{\langle i}{}^k q_{j\rangle}{}^l, \\
H^{(3)}_{(k)ij}{}^{(k)kl} &=&
\frac{1}{2} q_{\langle i}{}^k q_{j\rangle}{}^l.
\end{eqnarray*}

One can check that the symmetrizer so defined satisfies $H
(i|\omega|_h\PP) + (i|\omega|_h\PP)^{*} H =0$ and so trivially belongs
to $S^0_{1,0}$.  This symmetrizer $H(t,x,u,\tilde \omega)$ is bounded
for $b\neq 1$, $c>0$, and smooth in all its arguments. The case $b=1$
and $c=2$ can be computed in the same way described above, and the
same conclusion holds. Then, for these two cases the system
(\ref{eq:pseudo1})-(\ref{eq:pseudo3}) is strongly hyperbolic.\qed

Two further generalizations are immediate. The first one involves the
Hamiltonian constraint. Suppose the term $a h_{ij}$ times the
Hamiltonian constraint (\ref{eq:HC}) is added to Eq. (\ref{eq:SNd2}).
Here $a$ is any real constant.  What are the eigenvalues and
eigenvectors of the resulting principal symbol?  The result is, as one
expects, that the only change is in the scalar variable part
$\PP^{(1)}$.  The eigenvalues are
\[
\tilde \lambda^{(1)}_1 = \pm \sqrt{1+2a (2-c)}, \quad
\tilde \lambda^{(1)}_2 = \pm \sqrt{b},\quad
\tilde \lambda^{(1)}_3 = 0,
\]
each one having multiplicity 1, and $\lambda^{(1)}_1$ is real provided
$a (2-c) \geq -(1/2)$. The corresponding eigenvectors are \goodbreak
\[
\hat u^{(1)}_{\lambda_1} = \left[
\begin{array}{c}
2[(b-c+1) \tilde \omega_i \tilde \omega_j  + (1-b) q_{ij}/2] \\
\mp \lambda [(b-c+1) \tilde \omega_i \tilde \omega_j + (1-b) q_{ij}/2] \\
(2-c)b \tilde \omega_i
\end{array} \right]
\]
\[
+a(2-c) \left[
\begin{array}{c}
 2 (\tilde \omega_i \tilde \omega_j + q_{ij}) \\
\mp \lambda  (\tilde \omega_i \tilde \omega_j + q_{ij}) \\
(2c-1) \tilde \omega_i
\end{array} \right],
\]
\goodbreak
\[
\hat u^{(1)}_{\lambda_3} = \left[
\begin{array}{c}
2 \tilde \omega_i \tilde \omega_j \\
 0 \\
(1+b) \tilde \omega_i
\end{array} \right]
+a \left[
\begin{array}{c}
2[ (3-2b) \tilde \omega_i \tilde \omega_j + b q_{ij}] \\
0 \\
3 \tilde \omega_i
\end{array}\right],
\]
where $\lambda = \sqrt{1 + 2a(2-c)}$. The eigenvector
$\hat u^{(2)}_{\lambda_2}$ does not change. The conclusion is summarized
below.
\begin{corollary}
  Consider Eqs. (\ref{eq:SNd1})-(\ref{eq:SNd3}) and assume the
  hypothesis of Theorem \ref{T:BSSNsth}. Assume $b>0$ and $c>0$.
  Assume that a term $a h_{ij}$ times the Hamiltonian constraint
  (\ref{eq:HC}) is added to Eq. (\ref{eq:SNd2}), where $a$ is a real
  constant satisfying $a(2-c)>-1/2$. Then the resulting principal
  symbol, as defined in this Section, has real eigenvalues and a
  complete set of linearly independent eigenvectors.
\end{corollary}

The second generalization involves the transformation $F_i = f_i +
d\,\partial_i(\ln \h)$, where $d$ is any real constant.  That is,
instead of defining the BSSN-type system with the variable $f_i$,
define it with $F_i$.  The new evolution equations have a different
principal symbol from Eqs. (\ref{eq:SNd1})-(\ref{eq:SNd3}) of the
BSSN-type equations. However, one can check that it does not change
the hyperbolicity of the system. Indeed, it modifies only the
$\PP^{(1)}$ part of the principal symbol. Its eigenvalues remain the
same, namely, $\tilde \lambda^{(1)}_1 = \pm 1$, $\tilde \lambda^{(1)}_2
= \pm \sqrt{b}$, and $\tilde \lambda^{(1)}_3 = 0$. The associated
eigenvectors are now given by
\[
\hat u^{(1)}_{\lambda_1} = \left[
\begin{array}{c}
2[(b-c+1)\tilde \omega_i \tilde \omega_j + (1-b)(q_{ij}/2) ]\\
\mp[(b-c+1)\tilde \omega_i \tilde \omega_j + (1-b) (q_{ij}/2)] \\
(2-c) (b+d) \tilde \omega_i 
\end{array} \right],
\]
\[
\hat u^{(1)}_{\lambda_2} = \left[
\begin{array}{c}
2 \tilde \omega_i \tilde \omega_j \\
\mp \sqrt{b} \,\tilde \omega_i \tilde \omega_j \\
(1+d)\tilde \omega_i 
\end{array} \right], \quad
\hat u^{(1)}_{\lambda_3} = \left[
\begin{array}{c}
2 \tilde \omega_i \tilde \omega_j \\
0 \\
(1+b+d) \tilde \omega_i
\end{array} \right].
\]
Therefore, the hyperbolicity of the BSSN-type equations is not changed
by this transformation.

\section{Discussion}
\label{S:D}

The first order pseudodifferential reduction performed in the space
derivatives is here the main tool used to study the hyperbolicity of
the BSSN-type systems. This technique is widely used in
pseudodifferential analysis. It does not increase the number of
equations, so there are no new constraints added to the system. It
emphasizes that well posedness essentially captures the absence of
divergent behavior in the high frequency limit of the solutions of a
given system.  This tool is applied to Eqs.
(\ref{eq:SNd1})-(\ref{eq:SNd3}), which have derivatives of first order
in time and second order in space. They are obtained from the ADM
equations by densitizing the lapse function and introducing the three
connection variables $f_i$.  Its evolution equation is obtained by
adding the momentum constraint to an identity from commuting
derivatives. The positive-density lapse function and the spacelike
shift vector are arbitrary given functions. There are free parameters
given by the exponent in the densitized lapse and the factor in the
addition of the momentum constraint. The resulting first order
pseudodifferential system is strongly hyperbolic for some values of
the free parameters.  (See Theorem \ref{T:BSSNsth}.)

The introduction of $f_i$ as a new variable is inspired by the
variable $\tilde \Gamma^i$ of the BSSN system, defined by Eq. (21) in
\cite{tBsS98}, and in the study of the linearized ADM evolution
equations given in \cite{hKoO02}. This variable is the crucial step
that allows us to introduce the momentum constraint into the system.
These two things, in turn, produce the result that the vector variable
eigenvectors $\hat u^{(2)}_{\lambda}$ do span their eigenspace.  This
is the key feature that converts the weakly hyperbolic densitized ADM
system into a strongly hyperbolic one. This property does not change
when a term of the form $d \, \partial_i (\ln \h)$ is added to the
system.  Then, both results suggest why the BSSN system is preferred
to the ADM equations for numerical analysis. This conclusion agrees
with a previous result in \cite{oS-etal02}, where the hyperbolicity of
the BSSN system was also studied.  It is shown there that a
differential reduction to first order in time and space derivatives,
together with a densitization of the lapse, produce a strongly
hyperbolic system. The results in the present work are also consistent
with numerical studies on the evolution equations presented in
\cite{cBjM92,cB-etal95,aA-etal99}. For further developments on these
systems, see \cite{cB-etal02,cB-etal03}.

Finally, the role of the Hamiltonian constraint, when added to the ADM
and BSSN-type evolution equations, is studied. In the case where the
density lapse exponent $b\neq 1$ it does not affect the hyperbolicity
properties of the two systems. That is, densitized ADM evolution
equations remain weakly hyperbolic, and the BSSN-type system remain
strongly hyperbolic. In the case $b=1$ the addition of the Hamiltonian
constraint in both systems prevents the two eigenvectors of the scalar
variable block from collapsing onto each other. This keeps the
BSSN-type equation strongly hyperbolic even in the case $b=1$, but is
not enough to change the weakly hyperbolic character of the densitized
ADM evolution equations.

\begin{acknowledgments}
  We thank Olivier Sarbach for reading the manuscript and suggesting
  improvements. G.N. Also thanks Bruce Driver for discussions on
  pseudodifferential calculus. This work was partially supported by
  CONICET, and SeCyT, UNC.
\end{acknowledgments}

\appendix

\section{Essentials of pseudodifferential operators}
\label{e}

\subsection{Introduction}

Pseudodifferential operators are a generalization of differential
operators that make use of Fourier theory. The idea is to think of a
differential operator acting upon a function as the inverse Fourier
transform of a polynomial in the Fourier variable times the Fourier
transform of the function. This integral representation leads to a
generalization of differential operators, which correspond to
functions other than polynomials in the Fourier variable, as long as
the integral converges.

In other words, given a smooth complex valued function $\un p(x,\w)$
from $\R^n\times \R^n$ with some asymptotic behavior at infinity,
associate with it an operator $p(x,\partial_x): \Stz \to \Stz$. Here
$\Stz$ is the Schwartz space, that is, the set of complex valued
smooth functions in $\R^n$, such that the function and every
derivative decay faster than any polynomial at infinity. The
association $\un p \to p$, that is, functions into differential
operators, is not unique. This is known to anyone acquainted with
quantum mechanics. Different maps from functions $\un p(x,\w)$ into
operators $p(x,\partial_x)$ give rise to different theories of
pseudodifferential calculus. Every generalization must coincide in the
following: The polynomial $\un p(x,\w) =\sum_{|\alpha|\leq m}
a_{\alpha}(x) (i\w)^{\alpha}$ where $\alpha$ is a multi-index in
$\R^n$ must be associated with the operator $p(x,\partial_x)
=\sum_{|\alpha|\leq m} a_{\alpha}(x)\partial_x^{\alpha}$, that is,
with a differential operator of order $m$. The map $\un p \to p$ used
in these notes is introduced in subsection \ref{S:pdo} of the
Appendix. It is the most used definition of pseudodifferential
operators in the literature, and the one most studied.

The Fourier transform is used to rewrite the differential operator
because it maps derivatives into multiplication, that is, $[\partial_x
u(x)]\hat{~}= i\w \hat u(\w)$. This property is used to solve constant
coefficient partial differential equations (PDEs) by transforming the
whole equation into an algebraic equation. This technique is not
useful on variable coefficient PDEs, because of the inverse property,
that is, $[x u(x)]\hat{~}= i\partial_{\w} \hat u(\w)$. For example,
one has $[\partial_x u +x u]\hat{~} = i [\partial_{\w}\hat u + \w \hat
u]$, and nothing has been simplified by the Fourier transform. That is
why one looks for other ways of rewriting differential operators. The
generalization to pseudodifferential operators is an additional
consequence.  Other transforms can be used to define different
generalizations of differential operators. For example, Mellin
transforms are used in \cite{Schulze91}.

The functions $\un p(x,\w)$ are called symbols. Differential operators
correspond to polynomial symbols in $\w$. They contain the main
equations from physics. Even strongly hyperbolic PDEs have polynomial
symbols. Why should one consider more general symbols? Because the
generalization is evident, and it has proved worth doing it. The
Atiyah and Singer index theorem is proved using pseudodifferential
operators with smooth symbols, which are more suitable for studying
homotopy invariants than polynomial symbols \cite{Taylor96b}.
Techniques to prove the well posedness of the Cauchy problem for a
strongly hyperbolic system require one to mollify polynomial symbols
into smooth nonpolynomial ones \cite{Taylor81}. The main application
in these notes is simple: to reduce a second order partial
differential equation to a first order system without adding new
characteristics into the system. This is done by introducing the
operator, a square root of the Laplacian, which is a first order
pseudodifferential, but not differential, operator. The main idea for
this type of reduction was introduced in \cite{aC62}.

How far should this generalization be carried? In other words, how is
the set of symbols that define the pseudodifferential operators
determined? The answer depends on which properties of differential
operators one wants to be preserved by the general operators, and
which additional properties one wants the latter to have. There are
different spaces of symbols defined in the literature. Essentially all
of them agree that the associated space of pseudodifferential
operators is closed under taking the inverse. The inverse of a
pseudodifferential operator is another pseudodifferential operator.
This statement is not true for differential operators. The algebra
developed in studying pseudodifferential operators is useful to
compute their inverses. This is important from a physical point of
view, because the behavior of solutions of PDEs can be inferred from
the inverse operator. One could even say that pseudodifferential
operators were created in the middle 1950s from the procedure to
compute parametrices to elliptic equations. [A parametrix is a
function that differs from a solution of the equation $p(u) =\delta$
by a smooth function, where $p$ is a differential operator, and
$\delta$ is Dirac's delta distribution.] To know the parametrix is
essentially the same as to have the inverse operator.

Spaces of pseudodifferential operators are usually defined to be
closed under composition and transpose, and to act on distribution
spaces and on Sobolev spaces. They can be invariant under
diffeomorphism, so they can be defined on a manifold. This definition
for pseudodifferential operators is not so simple as for differential
operators, because the latter are local operators and the former are
not. Pseudodifferential operators are pseudolocal. An operator, $p$,
acting on a distribution, $u$, is local if $p(u)$ is smooth in the
same set where $u$ is smooth.  Pseudolocal means that the set where
$p(u)$ is smooth includes the set where $u$ is smooth. This means that
$p$ could smooth out a nonsmoothness of $u$. Mollifiers are an example
of this kind of smoothing operator. They are integral operators, which
justifies the name of pseudolocal. Differential operators with smooth
coefficients are an example of local operators.  The proofs of all
these properties of pseudodifferential operators are essentially
algebraic calculations on the symbols. One could say that the main
practical advantage of pseudodifferential calculus is, precisely,
turning differential problems into algebraic ones.

\subsection{Function spaces}
\label{S:fs}

The study of the existence and uniqueness of solutions to PDEs, as
well as the qualitative behavior of these solutions, is at the core of
mathematical physics. Function spaces are the basic ground for
carrying on this study. The mathematical structure needed is that of
the Hilbert space, or Banach space, or Fr\'echet space, which are
complete vector spaces having, respectively, an inner product, a norm,
and a particular metric constructed with a family of seminorms. Every
Hilbert space is a Banach space, and every Banach space is a Fr\'echet
space. The main examples of Hilbert spaces are the space of square
integrable functions $L^2$, and the Sobolev spaces $H^k$, with $k$ a
positive integer, which consist of functions whose $k$ derivatives
belong to $L^2$. The Fourier transform makes it possible to extend
Sobolev spaces to real indices. This generalization in the idea of the
derivative is essentially the same as one uses to construct
pseudodifferential operators. Examples of Banach spaces are $L^p$,
spaces of $p$-power integrable functions, where $L^2$ is the
particular case $p=2$. The main examples of Fr\'echet spaces are
$C^{\infty}(\Omega)$, the set of smooth functions in any open set
$\Omega \subset \R^n$, with a particular metric on it (the case
$\Omega=\R^n$ is denoted $C^{\infty}$), the Schwartz space of smooth
functions of rapid decrease, and its dual as a Fr\'echet space, which
is a space of distributions.

This section presents only Sobolev spaces, first with non-negative
integer index, and the generalization to a real index. The Fourier
transform is needed to generalize the Sobolev spaces. Therefore,
Schwartz spaces are introduced to facilitate the definition of the
Fourier transform, and to extend it to $L^2$. The next section is
dedicated to introducing pseudodifferential operators.

Let $L^2$ be the vector space of complex valued, square integrable
functions on $\R^n$, that is, functions such that $\|u\| < \infty$,
where $\|u\| := \sqrt{(u,u)}$ and
\[
(u,v) := \int_{\R^n} \bar u(x) v(x) dx,
\]
with $\bar u$ the complex conjugate of $u$.  This set is a Hilbert
space, that is, a complete vector space with inner product, where the
inner product is given by $(~,~)$ and is complete with respect to the
associated norm $\|~\|$.

The Sobolev spaces $H^k$, for $k$ a non-negative integer, are the
elements of $L^2$ such that
\[
\|u\|_k^2 := \sum_{|\alpha|\leq k} \|\partial^{\alpha}u\|^2 < \infty,
\]
where $\alpha= (\alpha_1, \alpha_2, \dots, \alpha_n)$ is a
multi-index, and for every such multi-index $\partial^\alpha$ denotes
$\partial_1^{\alpha_1}\partial_2^{\alpha_2}\dots\partial_n^{\alpha_n}$
and $|\alpha|=\sum_{i=1}^n \alpha_i$. The inner product in $L^2$
defines an inner product in $H^k$ given by
\[
(u,v)_k := \sum_{|\alpha|\leq k} 
(\partial^{\alpha} u, \partial^{\alpha} v).
\]

Let $\Stz$ be the space of functions of rapid decrease, also called
the Schwartz space, that is, the set of complex valued, smooth
functions on $\R^n$, satisfying
\[
|u|_{k,\alpha} := \sup_{x\in \R^n} |(1+|x|^2)^{k/2} \partial^{\alpha} u| 
< \infty
\]
for every multi-index $\alpha$, and all $k\in \N$ natural, with $|x|$
the Euclidean length in $\R^n$. The Schwartz space is useful in
several contexts. It is the appropriate space to introduce the Fourier
transform. It is simple to check that the Fourier transform is well
defined on elements in that space, in other words, the integral
converges. It is also simple to check the main properties of the
transformed function. More important is that the Fourier transform is
an isomorphism between Schwartz spaces. As mentioned earlier, the
Schwartz space provided with an appropriate metric is an example of a
Fr\'echet space. Its dual space is the set of distributions, which
generalizes the usual concept of functions.

The Fourier transform of any function $u \in \Stz$
is given by
\[
\F[u](x) = \hat u (x) := \int_{\R^n} e^{-i x\cdot \w} u(\w) \dbar \w,
\]
where $\dbar \w = d\w/(2\pi)^{n/2}$, while $d\w$ and $x\cdot \w =
\delta_{ij}x^i\w^j$ are the Euclidean volume element and scalar product in
$\R^n$, respectively. The map $\F:\Stz \to \Stz$ is an
isomorphism. The inverse map is given by
\[
\F^{-1}[u](x)= \check u (x) := \int_{\R^n} e^{i x\cdot \w} u(\w) \dbar \w.
\]
An important property of the Fourier transform useful in PDE theory is
the following: $[\partial_x^{\alpha} u(x)]\,\hat{} = i^{|\alpha|}
\w^{\alpha} \hat u(\w)$, and $[x^{\alpha} u(x)]\,\hat{} = i^{|\alpha|}
\partial_\w^{\alpha}\hat u(\w)$, that is, it converts smoothness of
the function into decay properties of the transformed function, and
vice versa. The Fourier transform is extended to an isomorphism $\F:
L^2 \to L^2$, first proving Parseval's theorem, that is, $(u,v) = (\hat
u, \hat v)$ for all $u$, $v\in \Stz$ (which gives Plancherel's formula
for norms, $\|u\|=\|\hat u\|$, in the case that the norm comes from an
inner product, with $u=v$) and second recalling that $\Stz$ is dense
in $L^2$.

The definition of Sobolev spaces $H^s$ for $s$ real is based on
Parseval's theorem. First recall that every $u\in H^k$ with
non-negative integer $k$ satisfies $\partial^{\alpha} u \in L^2$
for $|\alpha| \leq k$, so Parseval's theorem implies $|\w|^{k} \hat
u(\w)\in L^2$. Second, notice that there exists a positive
constant $c$ such that $(1/c)\W \leq (1+ |\w|) \leq c \W$, where $\W =
(1+|\w|^2)^{1/2}$. Therefore, one arrives at the following definition.
The Sobolev space $H^s$ for any $s\in \R$ consists of locally
square integrable functions in $\R^n$ such that $\W^s \hat u \in
L^2$. This space is a Hilbert space with the inner product
\[
(u,v)_s := \int_{\R^n} \W^{2s} \bar{\hat u}(\w) \hat v(\w) d\w,
\]
and the associated norm is denoted by
\[
\|u\|_s^2 := \int_{\R^n} \W^{2s} |\hat u(\w)|^2 d\w.
\]
One can check that $H^{s} \subset H^{s'}$ whenever $s'\leq s$. Notice
that negative indices are allowed. The elements of those spaces are
distributions. Furthermore, the Hilbert space $H^{-s}$ is the dual of
$H^s$. Finally, two more spaces are needed later on, $H^{-\infty}:=
\cup_{s\in\R} H^s$ and $H^{\infty} := \cap_{s\in\R} H^s$. These spaces
are, with appropriate metrics on them, Fr\'echet spaces. A closer
picture of the kind of element these spaces may contain is given by
the following observations. The Sobolev embedding lemma implies that
$H^{\infty} \subset C^{\infty}$, while the opposite inclusion is not
true. Also notice that $\Stz \subset H^{\infty}$, and therefore
$H^{-\infty}\subset\Stz'$, so the elements of $H^{-\infty}$ are
tempered distributions.

\subsection{Pseudodifferential operators}
\label{S:pdo}

Let $S^m$, with $m\in \R$, be the set of complex valued smooth
functions $\un p(x,\w)$ from $\R^n\times \R^n$, such that
\begin{equation}
\label{symb}
|\partial^{\beta}_x\partial^{\alpha}_\w \un p(x,\w) | 
\leq C_{\alpha} \W^{m-|\alpha|},
\end{equation}
with $C_{\alpha}$ a constant depending on the multi-index $\alpha$,
and $\W = (1+|\w|^2)^{1/2}$. This is the space of functions whose
elements are associated with operators. It is called the space of
symbols, and its elements $\un p(x,\w)$ symbols. There is no
asymptotic behavior needed in the $x$ variable, because Fourier
integrals are thought to be carried out in the $\w$ variable. The
asymptotic behavior of this variable is related to the order of the
associated differential operator, as one can shortly see in the
definition of the map that associates functions $\un p(x,\w)$ with
operators $p(x,\partial_x)$.  One can check that $S^{m'} \subset S^m$
whenever $m'\leq m$. Two more spaces are needed later on, $S^{\infty}
:=\cup_{m\in\R} S^m$ and $S^{-\infty} :=\cap_{m\in\R}S^m$.

Given any $\un p(x,\w) \in S^m$, the associated operator
$p(x,\partial_x):\Stz \to\Stz$ is said to belong to $\psi^m$, and is
determined by
\begin{equation}
\label{pdo}
p(x,\partial_x)(u) = \int_{\R^n} e^{ix\cdot \w} 
\un p(x,\w) \hat u(\w) \dbar \w,
\end{equation}
for all $u \in \Stz$. The constant $m$ is called the order of the
operator. It is clear that $u\in \Stz$ implies $p(x,\partial_x)(u) \in
C^{\infty}$; however, the proof that $p(u)\in\Stz$ is more involved.
One has to show that $p(u)$ and its derivatives decay faster than any
polynomial in $x$. The idea is to multiply Eq. (\ref{pdo}) by
$x^{\alpha}$ and recall the relation $i^{|\alpha|}x^{\alpha}
e^{i\w\cdot x} =\partial_\w^{\alpha} e^{i\w\cdot x}$. Integration by
parts and the inequality (\ref{symb}) imply that the resulting
integral converges and is bounded in $x$. This gives the decay.

The polynomial symbols $\un p(x,\w) =\sum_{|\alpha|=0}^m
a_{\alpha}(x)(i\w)^{\alpha}$ with non-negative integer $m$ correspond
to differential operators of order $m$, $p(x,\partial_x)
=\sum_{|\alpha|=0}^m a_{\alpha}(x)\partial_x^{\alpha}$. An example of
a pseudodifferential operator that is not differential is given by the
symbol $\un p(\w) = \chi(\w) |\w|^k\sin[\ln(|\w|)]$, where $k$ is a
real constant and $\chi(\w)$ is a cut function at $|\w|=1/2$. That is
a smooth function that vanishes for $|\w| \leq 1/2$ and is identically
1 for $|\w|\geq 1$. The cut function is needed to have a smooth
function at $\w=0$. This symbol belongs to $S^k$. The function $\un
p(\w) =\chi(\w)\ln(|\w|)$ is not a symbol, because $|\un p(\w)| \leq
c_0\W^{\epsilon}$, for every $\epsilon >0$, but $|\partial_\w \un
p(\w)|\leq c_1 \W^{-1}$, and the change in the decay is bigger than 1,
which is the value of $|\alpha|$ in this case. Another useful example
to understand the symbol spaces is $\un p(\w) = \chi(\w)
|\w|^k\ln(|\w|)$, with $k$ a real constant. This function is not a
symbol for $k$ natural or zero, for the same reason as in the previous
example.  However, it is a symbol for the remaining cases, belonging to
$S^{k+\epsilon}$, for every $\epsilon >0$.

A very useful operator is $\Lambda^s:\Stz\to\Stz$ given by
\[
\Lambda^s(u) := \int_{\R^n} e^{i\w\cdot x}
\W^s \hat u(\w) \dbar \w,
\]
where $s$ is any real constant. This is a pseudodifferential operator
that is not differential. Its symbol is $\un \Lambda^s = \W^s$, which
belongs to $S^s$, and then one says $\Lambda^s\in\psi^s$. It is
usually denoted as $\Lambda^s = (1- \Delta)^{s/2}$. It can be extended
to Sobolev spaces, that is, to an operator $\Lambda^s: H^{s} \to L^2$.
This is done by noticing the bound $\|\Lambda^s(u)\| =\|u\|_s$ for all
$u\in\Stz$, and recalling that $\Stz$ is a dense subset of $L^2$. This
operator gives a picture of what is meant by an $s$ derivative, for
$s$ real. One can also rewrite the definition of $H^s$, saying that
$u\in H^s$ if and only if $\Lambda^s(u)\in L^2$.

Pseudodifferential operators can be extended to operators acting on
Sobolev spaces. Given $p\in \psi^m$, it defines an operator
$p(x,\partial_x) : H^{s+m} \to H^s$. This is the reason to call $m$
the order of the operator. The main idea of the proof is again to
translate the basic estimate (\ref{symb}) in the symbol into an
$L^2$-type estimate for the operator, and then use the density of
$\Stz$ in $L^2$. The translation is more complicated for a general
pseudodifferential operator than for $\Lambda^s$, because symbols can
depend on $x$. Intermediate steps are needed, involving estimates on
an integral representation of the symbol, called the kernel of the
pseudodifferential operator. Pseudodifferential operators can also be
extended to act on distribution spaces $\Stz'$, the dual of Schwartz
spaces $\Stz$.

An operator $p:H^{-\infty}\to H^{-\infty}$ is called a smoothing
operator if $p(H^{-\infty})\subset C^{\infty}$. That means $p(u)$ is
smooth regardless of $u$ being smooth. One can check that a
pseudodifferential operator whose symbol belongs to $S^{-\infty}$ is a
smoothing operator. For example, $\un p(\w) = e^{-|\w|^2}\in
S^{-\infty}$. However, not every smoothing operator is
pseudodifferential. For example, $\un p(\w) = \rho(\w)$, with $\rho\in
H^{s}$ for some $s$ and having compact support, is a smoothing
operator which is not pseudodifferential unless $\rho$ is smooth.
Friedrichs' mollifiers, $J_{\epsilon}$ for $\epsilon \in (0,1]$, are a
useful family of smoothing operators, which satisfy
$J_{\epsilon}(u)\to u$ in the $L^2$ sense, in the limit $\epsilon \to
0$, for each $u \in L^2$.

Consider one more example, the operator $\lambda:\Stz \to \Stz$ given
by
\[
\lambda(u) := \int_{\R^n} e^{i\w\cdot x}
i|\w| \chi(\w)\hat u(\w) \dbar \w,
\]
where $\chi(\w)$ is again a cut function at $|\w|=1/2$. The symbol is
$\un \lambda(\w) = i|\w|\chi(\w)$. The cut function $\chi$ makes $\un
\lambda$ smooth at $\w=0$. The operator without the cut function is
$\ell: \Stz\to L^2$ given by
\[
\ell(u) := \int_{\R^n} e^{i\w\cdot x}
i|\w| \hat u(\w) \dbar \w.
\]
Its symbol $\un \ell(\w) = i|\w|$ does not belong to any $S^m$ because
it is not smooth at $\w=0$. Both operators $\lambda$, $\ell$ can be
extended to maps $H^1 \to L^2$. What is more important, their
extensions are essentially the same, because they differ in a
smoothing, although not pseudodifferential, operator.

The asymptotic expansion of symbols is maybe the most useful notion
related to pseudodifferential calculus. Consider a decreasing sequence
$\{m_j\}_{j=1}^{\infty}$, with $\lim_{j\to \infty} m_j = -\infty$. Let
$\{\un p_j\}_{j=1}^{\infty}$ be a sequence of symbols $\un p_j(x,\w)\in
S^{m_j}$. Assume that these symbols are asymptotically homogeneous in
$\w$ of degree $m_j$, that is, they satisfy $\un p_j(x,t\w) = t^{m_j} \un
p_j(x,\w)$ for $|\w| \geq 1$. Then, a symbol $\un p \in S^{m_1}$ has the
asymptotic expansion $\sum_j \un p_j$ if and only if
\begin{equation}
\label{asymp}
\left(\un p - \sum_{j=1}^{k} \un p_j \right)\in S^{m_{(k+1)}},
\quad \forall \; k\geq 1,
\end{equation}
and it is denoted by $\un p \sim \sum_j \un p_j$. The first order term
in the expansion, $\un p_{1}$, is called the principal symbol.
Notice that $m_j$ are real constants, not necessarily integers. Every
asymptotic expansion defines a symbol, that is, every function of the
form $\sum_j\un p_j$ belongs to some symbol space $S^{m_1}$. However,
not every symbol $\un p \in S^m$ has an asymptotic expansion. Consider
the example $\un p(\w) =\chi(\w)|\w|^{1/2}\ln(|\w|)$.  The set of symbols
that admit an asymptotic expansion of the form (\ref{asymp}) is called
classical, it is denoted by $\Scl^m$, and the corresponding operators
are said to belong to $\psicl^m$. One then has $\Scl^m \subset S^m$.
Notice that if two symbols $\un p$ and $\un q$ have the same
asymptotic expansion $\sum_j \un p_j$, then they differ in a
pseudodifferential smoothing operator, because
\[
\un p - \un q = \left( \un p - \sum_{j=1}^{k} \un p_j\right)
-\left( \un q - \sum_{j=1}^{k} \un p_j\right) \in S^{m_{(k+1)}},
\]
for all $k$, and $\lim_{k\to\infty} m_j = -\infty$, so $(\un p -\un
q)\in S^{-\infty}$. This is the precise meaning for the rough sentence,
``what really matters is the asymptotic expansion.''

There is in the literature a more general concept of asymptotic
expansion. It does not require that the $\un p_{j}$ to be
asymptotically homogeneous. We do not consider this generalization in
these notes.

Most of the calculus of pseudodifferential operators consists of
performing calculations with the highest order term in the asymptotic
expansion and keeping careful track of the lower order terms. The
symbol of a product of pseudodifferential operators is not the
product of the individual symbols. Moreover, the former is difficult
to compute. However, an asymptotic expansion can be explicitly written
for classical symbols, and one can check that the principal symbol of
the product is equal to the product of the individual principal
symbols. More precisely, given $p\in\psicl^r$ and $q\in\psicl^s$, then
the product is a well defined operator $pq\in\psicl^{r+s}$ and the
asymptotic expansion of its symbol is
\[
\un{pq} \sim \sum_{|\alpha|\geq 0} \frac{1}{i^{|\alpha|} \alpha!}
\left[\partial^{\alpha}_\w \un p(x,\w)\right]
\left[ \partial^{\alpha}_x\un q(x,\w)\right].
\]
Notice that the first term in the asymptotic expansion of a commutator
$[p,q]=pq-qp$, that is its principal symbol, is precisely $1/i$ times
the Poisson bracket of their respective symbols, $ \{\un p,\un
q\}=\sum_j \left(\partial_{\w^j}\un p\partial_{x^j} \un q
  -\partial_{x^j}\un p \partial_{\w^j}\un q\right)$.  

Similarly, the symbol of the adjoint pseudodifferential operator is
not the adjoint of the original symbol. However, this is true for the
principal symbols. The proof is based in an asymptotic expansion of
the following equation:
\[ 
\un{(p^{*})}(x,\w) = \int\int_{\R^n} e^{-i(x-x')\cdot(\w-\w')}
(\un p)^{*}(x',\w')\, \dbar x \dbar \w.
\]

There are three main generalizations of the theory of
pseudodifferential operators present in the literature. First, the
operators act on vector valued functions instead of on scalar
functions.  While this is straightforward, the other generalizations
are more involved. Second, the space of symbols is enlarged, first
done in \cite{lH67}. It is denoted as $S^m_{\rho\delta}$, and its
elements satisfy $|\partial^{\beta}_x\partial^{\alpha}_\w \un
p(x,\w)|\leq C_{\alpha,\beta} \W^{m-\rho |\alpha|+\delta|\beta|}$,
with $C_{\alpha,\beta}$ a constant depending on the multi-indices
$\alpha$ and $\beta$. The extra indices have been tuned to balance two
opposite tendencies; on the one hand, to preserve some properties of
differential operators; on the other, hand to maximize the amount of
new objects in the generalization. These symbol spaces contain
functions like $\un p(x,\w) = \W^{a(x)}$, which belongs to
$S^m_{1,\delta}$, where $\delta>0$ and $m = \max_{x\in \R^{n}} a(x)$.
Third, the domain of the functions $\un p(x,\w)$ is changed from
$\R^n\times\R^n$ to $\Omega\times\R^n$, with $\Omega\subset\R^n$ any
open set. A consequence in the change of the domain is that
$p:C^{\infty}_0(\Omega)\to C^{\infty}(\Omega)$, so the domain and
range of $p$ are not the same, which makes it more difficult to define
the product of pseudodifferential operators. These notes are intended
to be applied to hyperbolic PDEs on $\R^n$, which are going to be
converted to pseudodifferential operators in $S^1$, so there is no
need to consider the last two generalizations.

\subsection{Further reading}

There is no main reference followed in these notes; however, a good
place to start is \cite{Taylor81}. Notice that the notation is not
precisely the one in that reference. The introduction is good, and the
definitions are clear. The proofs are difficult to follow. More
extended proofs can be found in \cite{Petersen83}, together with some
historical remarks. The whole subject is clearly written in
\cite{lN73}. It is not the most general theory of pseudodifferential
operators, but it is close to these notes. A slightly different
approach can be found in \cite{Friedrichs68}, and detailed calculations
to find parametrices are given in \cite{Treves80}. The introduction of
\cite{Kumano-go81} is very instructive. The first order reduction
using $\Lambda$ is due to Calder\'on in \cite{aC62}, and a clear
summary of this reduction is given in \cite{lN73}.

The field of pseudodifferential operators grew out of a special class
of integral operators called singular integral operators. Mikhlin in
1936 and Calder\'on and Zygmund in the begining of 1950s carried out the
first investigations. The field started to develop really fast after a
suggestion by Peter Lax in 1963 \cite{Lax63}, who introduced the
Fourier transform to represent singular integral operators in a
different way. Finally, the work of Kohn and Nirenberg \cite{jKlN65}
presented the pseudodifferential operators as they are known today,
and they proved their main properties. They showed that singular
integral operators are the particular case of pseudodifferential
operators of order zero. Further enlargements of the theory were due
to Lars H\"ormander \cite{lH65,lH67}.


\end{document}